\documentclass[11pt,aps,a4paper,eqsecnum,amsmath,amssymb,nofootinbib,longbibliography,notitlepage]{revtex4-1}

\pdfoutput=1

\usepackage{color}
\usepackage{amsfonts}
\usepackage{graphicx}
\usepackage{amsmath}
\usepackage[caption=false]{subfig}
\usepackage{tikz}

\begin{document}

\preprint{}
\title{%
  Condensates of interacting non-Abelian $SO(5)_{N_f}$ anyons }

\author{Daniel Borcherding}
\author{Holger Frahm}
\affiliation{%
  Institut f\"ur Theoretische Physik, Leibniz Universit\"at Hannover,
  Appelstra\ss{}e 2, 30167 Hannover, Germany}

\date{\today}

\begin{abstract}
  Starting from a one-dimensional model of relativistic fermions with
  $SO(5)$ spin and $U(N_f)$ flavour degrees of freedom we study the
  condensation of $SO(5)_{N_f}$ anyons. In the low-energy limit the
  quasi-particles in the spin sector of this model are found to be
  massive solitons forming multiplets in the $SO(5)$ vector or spinor
  representations. The solitons carry internal degrees of freedom
  which are identified as $SO(5)_{N_f}$ anyons. By controlling the
  external magnetic fields the transitions from a dilute gas of free
  anyons to various collective states of interacting ones are
  observed.  We identify the generalized parafermionic cosets
  describing these collective states and propose a low temperature
  phase diagram for the anyonic modes.
\end{abstract}

\maketitle


\section{Introduction}

The fractionalized quasi-particle excitations of topological states of
matter have attracted a lot of attention in the recent years. A
particulary interesting class of these quasi-particles are so-called
non-Abelian anyons.  Their remarkable exchange statistics makes them a
resource for decoherence-free quantum computing \cite{Kita03}
which has further driven the search for physical
realizations. Candidate systems are the topologically ordered phases
of two-dimensional quantum matter such as the fractional quantum Hall
states or $p+\mathrm{i}p$ superconductors where non-Abelian anyons may
appear as zero-energy degrees of freedom of gapped excitations in the
bulk \cite{MoRe91,ReRe99,ReGr00}.

Mathematically, anyons are objects in a braided tensor category. In
this description they are characterized by their braiding and fusion
properties.  These completely determine the physics of a dilute anyon
gas and quantum computing operations can be realized based on the
braiding of the anyonic quasi-particles \cite{NSSF08}.  The fusion
rules on the other hand determine the Hilbert space of a many-anyon
system as well as the possible local interactions between pairs of
anyons \cite{FTLT07}.  The presence of the latter lifts the degeneracy of the
zero-energy modes and leads to the anomalous collective behaviour of
systems with a finite density of anyons, e.g.\ when they condense at
the boundaries between phases of different topological order.  This
can be exploited to stabilize topological quantum memories
\cite{Brown.etal16}.

The properties of interacting anyons forming a high density condensate
on the edge of the topologically ordered phase of a two dimensionsal
quantum system have been studied in various effective lattice models
\cite{FTLT07,GATH13,Finch.etal14,FiFF14,BrFF16,VeJS17,FiFF18}.
Combining numerical methods with insights from exactly solvable models
and conformal field theory important insights into the collective
behaviour of different types of non-Abelian anyons have been obtained.
Unfortunately, these lattice models do not allow to tune the anyon
density.  To study the transition between the low density phase of
'bare' anyons and the collective state realized at high anyon
densities one can follow the approach of
\cite{Tsve14a,BoFr18,BoFr18a}:
at sufficiently low temperatures the Hamiltonian of an integrable
one-dimensional model of fermions carrying $SU(k)$ spin and $U(N_f)$
flavour indices with a particular current-current interaction can be
separated into commuting parts describing the fractionalized charge,
spin and flavour degrees of freedom separately.  Concentrating on the
spin sector the elementary excitations are found to be massive
solitons forming multiplets in fundamental representations of $SU(k)$
and bound states thereof.  Residing on these solitons are
$SU(k)_{N_f}$ anyons.  Therefore the density of the anyonic degrees of
freedom can be controlled together with that of the solitons by the
variation of the magnetic fields.  This allows, by solving the
thermodynamic Bethe ansatz equations for different external fields, to
study the condensation of $SU(k)_{N_f}$ anyons in detail.

In the present paper we extend this approach to fermions with an
$SO(5)$ spin degree of freedom.\footnote{%
  $SO(5)$ symmetric electron models have been constructed e.g.\ in
  Refs.~\cite{ScZH98,FrSt01}. In the present paper, however, we do not
  discuss the origin of this and the additional flavour degrees of
  freedom but rather concentrate on the possible existence of
  $SO(5)_{N_f}$ anyons in such models and their signature in the
  thermodynamical properties.}
Specifically we consider a model defined by Hamiltonian densities
describing relativistic chiral fermions in external magnetic fields
$h_i$ ($i=1,2$) perturbed by an anisotropic spin-spin interaction
\begin{align}
  \begin{split}
    \label{so5_NfModel}
    \mathcal{H}&=
    -\mathrm{i}\bar{\psi}_{f}\gamma_\mu \partial^{\mu}\psi_f
    -\sum_{f=1}^{N_f}h_i\,\bar{\psi}_f\gamma_0H^i_c\psi_f
    +\mathcal{H}_{\text{int}}\,,\\
    \mathcal{H}_{\text{int}}&
    =\lambda_{\parallel}\sum_{i=1}^{2}(\bar{\psi}_f\gamma_\mu H^{i}_c\psi_f)^2
    +\lambda_{\perp}\sum_{\alpha>0}\frac{|\alpha|^2}{2}
      (\bar{\psi}_f\gamma_\mu E^{\alpha}\psi_f) (\bar{\psi}_{f'}\gamma_\mu E^{-\alpha}\psi_{f'})\,,\\
  \end{split}
\end{align}
where $\psi_{fa}$ are Dirac spinors with $U(N_f)$ `flavour' indices
$f,f'=1,\dots,N_f$ and $SO(5)$ `spin' indices $a=1,\dots,5$ (the
latter are suppressed in (\ref{so5_NfModel})). $H^i_{c}$ ($i=1,2$) are
the generators of the $SO(5)$ Cartan subalgebra while the $SO(5)$
ladder operator for a root $\alpha$ in the Cartan-Weyl basis is
denoted by $E^\alpha$. Moreover, the $\gamma_\mu$ ($\mu=1,2$) are
Dirac matrices and $\bar{\psi}_f=\gamma_0\psi^\dagger_f$.

Similar as in the models for fermions carrying $SU(k)$ spin mentioned
above we find that the excitations in the spin sector of
(\ref{so5_NfModel}) are massive solitons. Here they form multiplets in
the $SO(5)$ vector and spinor representations and, in addition, carry
an internal degree of freedom which we identify as non-Abelian
$SO(5)_{N_f}$ anyons.  The density of solitons can be controlled by
the external fields coupling to the $SO(5)$ Cartan generators.  For
sufficient large magnetic fields solitons form a condensate described
by a $U(1)$ Gaussian model.  The condensation of the solitons is
accompanied by the formation of collective states for the anyon
degrees of freedom which are found to be described by generalized
parafermionic conformal field theories.  We note that this complements
previous results obtained for the high density collective states of
interacting $SO(5)_2$ anyons in lattice models \cite{FiFF14,FiFF18}.

\section{Bethe ansatz for a perturbed $\mathbf{SO(5)_{N_f}}$
  WZNW model}
\label{sec:SO5_IntFermions}
In the models of fermions with $SU(k)$ spin studied previously
\cite{Tsve14a,BoFr18,BoFr18a} conformal embedding has been used to
isolate the part of the Hamiltonian describing the collective
excitations in the spin sector \cite{FrMSbook96}.  Here, we rely
instead on the spectrum of the model obtained from the exact solution
of (\ref{so5_NfModel}):
the isotropic model, $\lambda_\parallel=\lambda_\perp$ has been solved
using the Bethe ansatz \cite{PoWi83,OgWi86,OgRW87}. In the limit
$N_f\to\infty$ the fermionic model is equivalent to the
$SO(5)\times SO(5)$ chiral model and its spectrum and magnetic
properties have been studied in Ref.~\cite{Naka86}.
For generic anisotropic choice of the coupling constants the
integrability of the Hamiltonian is based on a deformation of the
corresponding factorized scattering matrices
\cite{Bazhanov85,Jimb86,Bazhanov87,ReWi87}.  The low-energy excitations of
(\ref{so5_NfModel}) carry charge, flavour and spin degrees of freedom
(cf. \cite{OgRW87} for the isotropic case).  Here we are interested
solely in the $SO(5)$ spin degrees of freedom. By placing
$\mathcal{N}$ fermions into a box of length $L$ with periodic boundary
conditions the energy contribution of the spin excitations specified
by quantum numbers $N_1,\,N_2$ is
\begin{equation}
  \label{so5_energy}
  E=\frac{\mathcal{N}}{L}\sum_{\alpha=1}^{N_1}\sum_{\tau=\pm 1}\frac{\tau}{2\mathrm{i}}\ln\left(\frac{\sinh(\frac{\pi}{2p_0}(\lambda^{(1)}_\alpha+\tau/g-N_f\mathrm{i}))}{\sinh(\frac{\pi}{2p_0}(\lambda^{(1)}_\alpha+\tau/g+N_f\mathrm{i}))} \right)+N_1 H_1 +N_2 H_2 -\mathcal{N}\left(H_1+H_2\right)\,,
\end{equation}
where $g$, $p_0$ are functions of the coupling constants
$\lambda_{\parallel}$ and $\lambda_\perp$ in
(\ref{so5_NfModel}). $H_1$ and $H_2$ are linear combinations of the
magnetic fields introduced above, i.e.\
$H_1\equiv{\alpha}^1\cdot\vec{h}$, $H_2\equiv{\alpha}^2\cdot\vec{h}$
with the simple roots ${\alpha}^j$, $j=1,2$, of $SO(5)$, see
Appendix~~\ref{app:SO5}. Also notice that the relativistic invariance
of the fermion model is broken by the choice of boundary conditions
but will be restored later by considering observables in the scaling
limit $L,\mathcal{N}\rightarrow \infty$ and $g\ll 1$ such that the
mass of the elementary excitations is small compared to the particle
density $\mathcal{N}/L$. The complex parameters $\lambda_\alpha^{(m)}$
with $\alpha=1,\dots,N_m$ ($m=1,2$) appearing in (\ref{so5_energy})
are so called Bethe roots solving the hierachy of Bethe equations
(cf. Refs. \cite{Resh85,Naka86} for the isotropic case)
\begin{equation}
  \label{so5_betheeq}
  \begin{aligned}
    \prod_{\tau=\pm 1}e_{N_f}(\lambda^{(1)}_\alpha+\tau/g)^{\mathcal{N}/2}&=\prod_{\beta \neq \alpha}^{N_1}e_2(\lambda^ {(1)}_\alpha - \lambda^{(1)}_{\beta})\prod_{\beta=1}^{N_2}e_{-1}(\lambda^{(1)}_\alpha-\lambda^{(2)}_\beta)\,,\quad \alpha=1,\ldots N_1\,,\\
    \prod_{\beta=1}^{N_1}e_{1}(\lambda^{(2)}_\alpha-\lambda^{(1)}_\beta)&=\prod_{\beta \neq \alpha}^{N_2}e_1(\lambda^{(2)}_\alpha-\lambda^{(2)}_\beta)\,,\quad \alpha=1,\ldots,N_2\,,
  \end{aligned}
\end{equation}
where
$e_k(x)=\sinh\left(\frac{\pi}{2p_0}(x+{\mathrm
    i}k)\right)/\sinh\left(\frac{\pi}{2p_0}(x-{\mathrm i}k)\right)$.

\begin{figure}[t]%
    \centering
    \captionsetup[subfigure]{singlelinecheck=on}
    \subfloat[weights]{{\includegraphics[width=7cm]{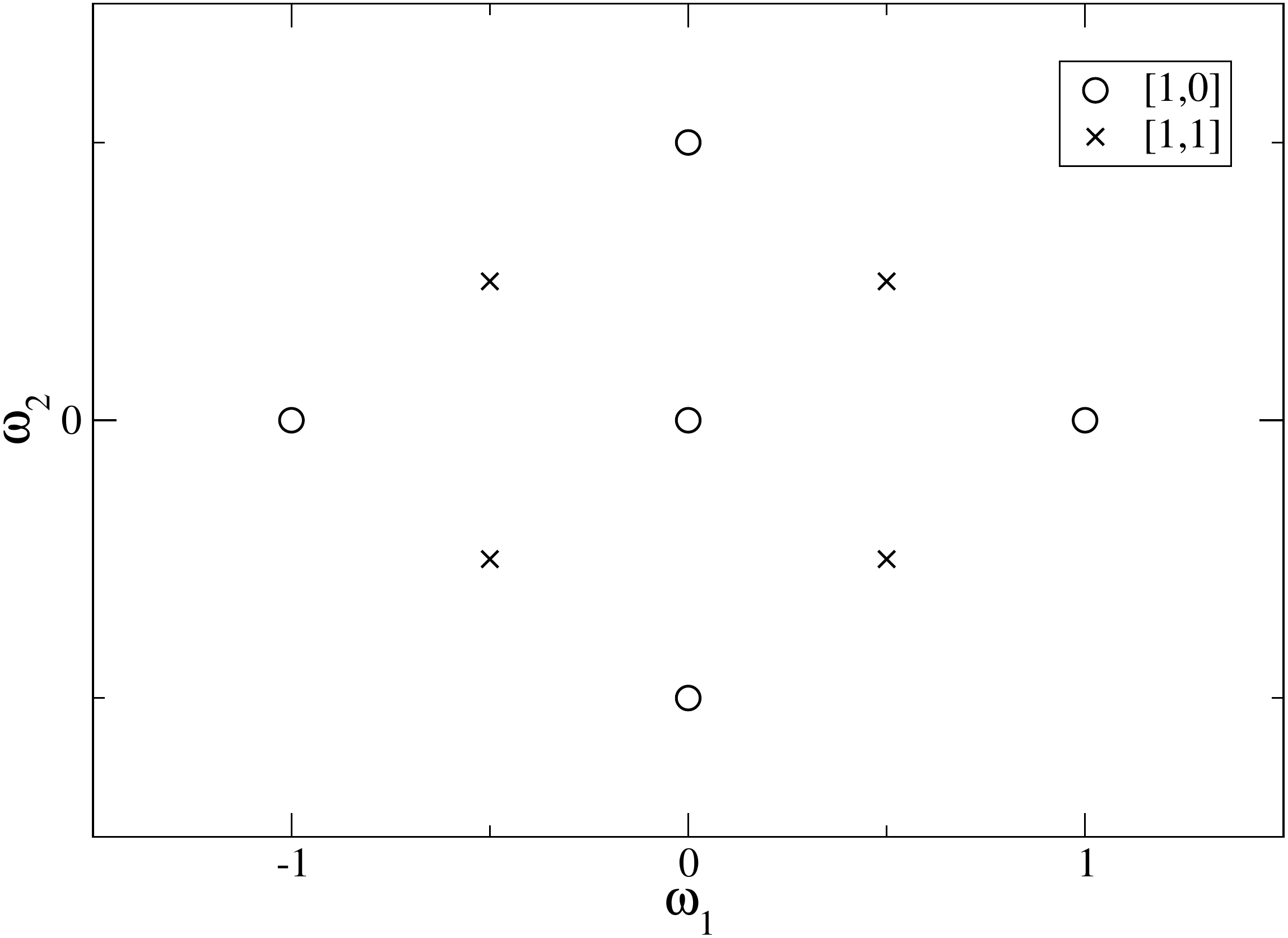} }}%
    \qquad
    \subfloat[charges]{{\includegraphics[width=7cm]{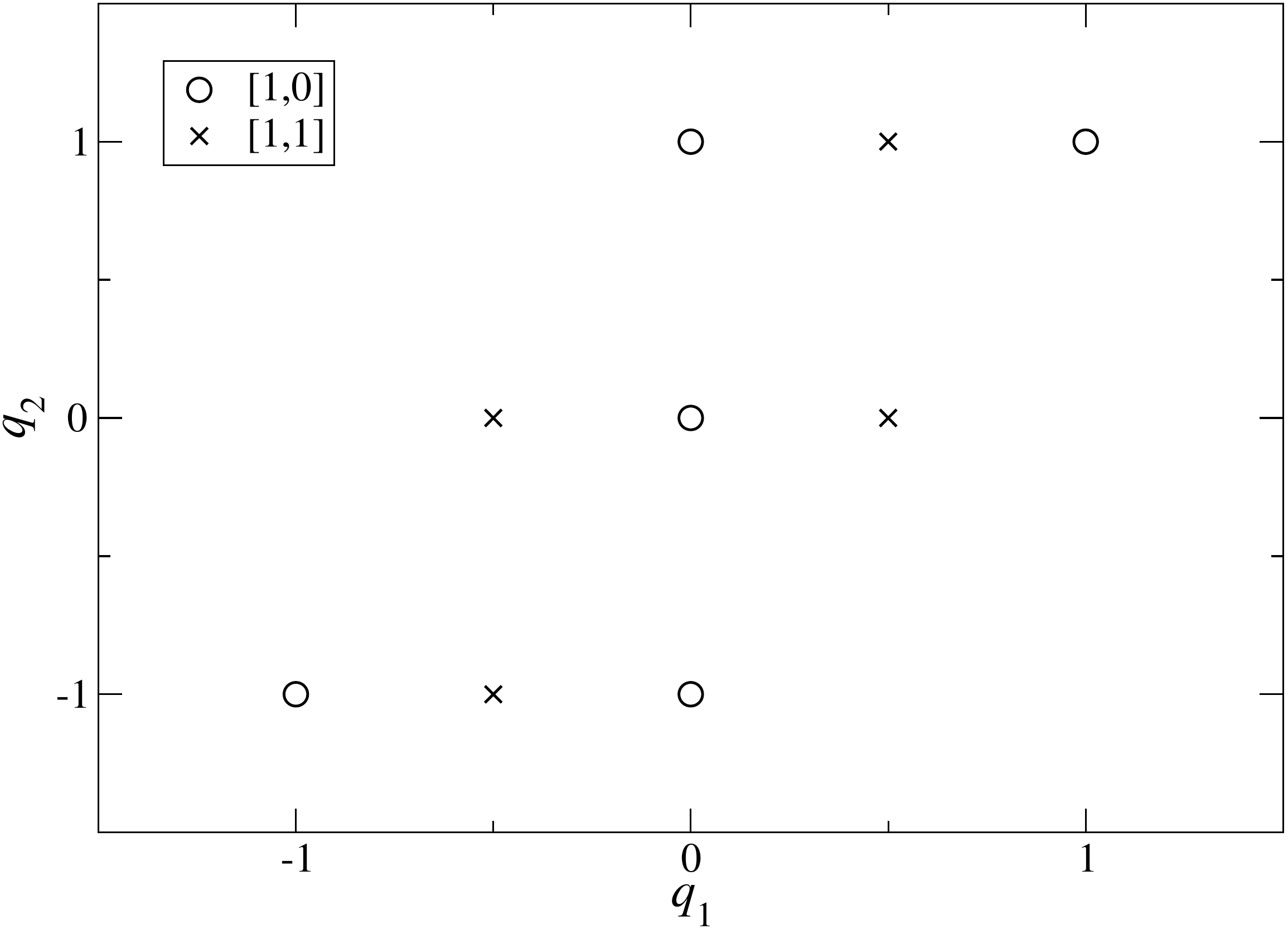} }}%
    \caption{Relationship between the weights (a) of the $[1,0]$
      (Dynkin labels $(1,0)$) and $[1,1]$ (Dynkin labels $(0,1)$)
      representation of ${SO}(5)$ and the charges (b) determined by
      (\ref{so5_charges}) with the projection of the magnetic fields
      $\vec{h}=(h_1,h_2)$ in (\ref{so5_NfModel}) on the directions of
      the simple roots, $H_1={\alpha^1}\cdot \vec{h}$,
      $H_2={\alpha^2}\cdot \vec{h}$.}%
    \label{fig:so5_weightscharges}%
\end{figure}

Based on equations (\ref{so5_betheeq}), (\ref{so5_energy}) the
thermodynamics of the model can be studied provided that the solutions
to the Bethe equations describing the eigenstates in the limit
$\mathcal{N}\to\infty$ are known.  Here we argue that the root
configurations corresponding to the ground state and excitations
relevant for the low-temperature behavior of (\ref{so5_NfModel}) can
be built based on a generalized string hypothesis, see e.g.\
Refs.~\cite{TaSu72,Martins91}: in the thermodynamic limit the Bethe
roots $\lambda_\alpha^{(m)}$ 
are grouped into $j$\textit{-strings} of length $n_j$ and with parity
$v_{n_j}\in\{\pm 1\}$
 \begin{equation}
    \label{so5_string}
    \begin{aligned}
	    \lambda^{(1)}_{j,\alpha,\ell}&=\lambda^{(1)}_{j,\alpha} + i\left(n_j+1-2\ell\right)+\frac{p_0}{2}(1-v^{(1)}_{N_f}v^{(1)}_{n_j}), \quad \quad \ell=1,\dots,n_j\,,\\
	    \lambda^{(2)}_{j,\alpha,\ell}&=\lambda^{(2)}_{j,\alpha} + \frac{i}{2}\left(n_j+1-2\ell\right)+\frac{p_0}{2}(1-v^{(2)}_{2N_f}v^{(2)}_{n_j}), \quad \quad \ell=1,\dots,n_j
	\end{aligned}
\end{equation}
with real centers $\lambda^{(m)}_{j,\alpha} \in \mathbb{R}$.  The allowed lengths and parities depend on the parameter $p_0$. To simplify the the discussion below we assume that $p_0=N_f+1/\nu$ with integer $\nu>2$ where only a few string configurations are relevant for the low-temperature thermodynamics
\begin{alignat}{3}
		&n^{(1)}_{j_{2,1}}= {j_{2,1}}\,, \quad 
		    &&v^{(1)}_{j_{2,1}}=1\,, \qquad &&1\leq j_{2,1}\leq N_f-1\,,\nonumber\\
		&n^{(2)}_{j_{2,2}}= {j_{2,2}}\,, \quad 
		    &&v^{(2)}_{j_{2,2}}=1\,, \qquad &&1\leq j_{2,2}\leq 2N_f-1\,,\nonumber\\
		&n^{(1)}_{\tilde{j}_{0,1}}=N_f(\nu-1)+1\,, \quad 
		    &&v^{(1)}_{\tilde{j}_{0,1}}=(-1)^{\nu}\,, \qquad &&\tilde{j}_{0,1} =N_f\,,\nonumber\\
		&n^{(1)}_{j_{0,1}}=N_f\,, \quad &&v^{(1)}_{j_{0,1}}=1\,, \qquad &&j_{0,1}=N_f+1\,,\nonumber\\
		&n^{(2)}_{j_{0,2}}=2N_f\,, \quad &&v^{(2)}_{j_{0,2}}=1\,, \qquad &&j_{0,2}=2N_f+1\,,\nonumber
	\end{alignat}
together with the string configuration for even $\nu$
\begin{equation}\nonumber
  n^{(2)}_{\tilde{j}_{0,2}}=2N_f(\nu/2-1)+1\,, \quad
  v^{(2)}_{\tilde{j}_{0,2}}=(-1)^{\nu/2}\,, \qquad \tilde{j}_{0,2} =2N_f\,.\nonumber\\
\end{equation}
Within the root density approach the Bethe equations are rewritten as
coupled integral equations for the densities of these strings
\cite{YaYa66b}. For vanishing external fields one finds that the Bethe
root configuration corresponding to the lowest energy state is
described by finite densities of $j_{0,m}$-strings on the levels
$m=1,2$. The elementary excitations above this ground state are of
three types: similar as for the discussion of the perturbed
${SU}(3)_{N_f}$ WZNW model the excitations corresponding to holes in
the distributions of $j_{0,m}$-strings on level $m=1,2$ are
solitons. From their coupling to the fields it is found that they
carry quantum numbers of the five-dimensional vector representation of
${SO}(5)$ with Young diagram $[1,0]$ and the four-dimensional spinor
representation $[1,1]$, respectively. Hence, we refer to solitons of
the first level as $[1,0]$-solitons and to solitons of the second
level as $[1,1]$-solitons. The excitations corresponding to
$j_{2,m}$-strings are denoted by auxiliary modes, while the
contributions of breather excitations are assumed to be negligible for
low temperatures.
The densities $\rho^{(m)}_j(\lambda)$ of these excitations (and $\rho^{h(m)}_j(\lambda)$ for the corresponding holes) satisfy the integral equations
    \begin{equation}
    \label{so5_densities2}
        \rho^{h(m)}_k(\lambda)=\rho^{(m)}_{0,k}(\lambda)-\sum_{l=1}^{2}\sum_jB^{(m,l)}_{kj}\ast \rho^{(l)}_j \qquad m=1,2\,,
    \end{equation}
see Appendix~\ref{so5_app:onTBA}. As mentioned above relativistic invariance is restored in the scaling limit $g\ll 1$ where the solitons are massive particles with bare densities $\rho^{(m)}_{0,j_{0,m}}$ and bare energies $\epsilon^{(m)}_{0,j_{0,m}}$
\begin{align}
\label{so5_baredensity1}
        \rho^{(m)}_{0,j_{0,m}}(\lambda)\stackrel{g\ll 1}{=}\begin{cases}
            \frac{2\sqrt{3}M_0}{6}\cosh(\pi \lambda/3) \quad &\text{if }m=1\,,\\
            \frac{2M_0}{6}\cosh(\pi \lambda/3) \quad &\text{if }m=2\,,
        \end{cases}
\end{align}
\begin{align}
        \label{so5_energysol1}
        \epsilon^{(m)}_{0,j_{0,m}}(\lambda)\stackrel{g\ll 1}{=}\begin{cases}
            2\sqrt{3}M_0\cosh(\pi \lambda/3)-zH_1-zH_2 \quad &\text{if }m=1\,,\\
            2M_0\cosh(\pi \lambda/3)-\frac{z}{2}H_1-zH_2 \quad &\text{if }m=2\,,
        \end{cases}
\end{align}
where $z=(1+N_f\nu)$. The prefactors $2\sqrt{3}M_0$ and $2M_0$ with
$M_0\equiv e^{-\pi/{3g}}$ are the masses of the $[1,0]$- and
$[1,1]$-solitons, respectively. Furthermore, the corresponding charges
can be read off from (\ref{so5_energysol1}): for a general excitation
with mass $M$ and bare energy $\epsilon_0(\lambda)$ its charges
$(q_1,q_2)$ are defined by
\begin{equation}
  \label{so5_charges}
  \epsilon_0(\lambda)= M\cosh\left(\frac{\pi\lambda}{3}\right)
  -z \left(\omega_1 h_1 + \omega_2 h_2\right)
  =M\cosh\left(\frac{\pi\lambda}{3}\right)-z\left( q_1 H_1 + q_2 H_2\right)\,,
\end{equation}
where $\omega_1,\,\omega_2$ are the components of a weight in a
${SO}(5)$ representation. Consequently, $[1,0]$-solitons of type
$j_{0,1}$ carry the charge $(q_1,q_2)=(1,1)$, while $[1,1]$-solitons
of type $j_{0,2}$ carry the charge $(q_1,q_2)=(1/2,1)$. These charges
correspond to the highest weight states of the $[1,0]$ (vector) and
$[1,1]$ (spinor) representation of ${SO}(5)$. All possible charges of
$[1,0]$- and $[1,1]$-solitons are shown in
Figure~\ref{fig:so5_weightscharges}.

Similarly, the bare densities and energies of the $\tilde{j}_{0,m}$-strings are
\begin{align}
\label{so5_baredensity2}
        \rho^{(m)}_{0,\tilde{j}_{0,m}}(\lambda)\stackrel{g\ll 1}{=}\begin{cases}
            \frac{2\sqrt{3}M_0}{6}\cosh(\pi \lambda/3) \quad &\text{if }m=1\,,\\
            \frac{2M_0}{6}\cosh(\pi \lambda/3) \quad &\text{if }m=2\,,
        \end{cases}
\end{align}
\begin{align}
        \label{so5_energysol2}
        \epsilon^{(m)}_{0,\tilde{j}_{0,m}}(\lambda)\stackrel{g\ll 1}{=}\begin{cases}
            2\sqrt{3}M_0\cosh(\pi \lambda/3)-zH_2 \quad &\text{if }m=1\,,\\
            2M_0\cosh(\pi \lambda/3)-\frac{z}{2}H_1 \quad &\text{if }m=2\,.
        \end{cases}
\end{align}
The corresponding masses of these excitations coincide with the masses
of the $[1,0]$- and $[1,1]$-solitons, respectively. However, they
couple to these modes in a different way,
i.e. $\tilde{j}_{0,1}$-strings carry the charge $(0,1)$ and
$\tilde{j}_{0,2}$-strings the charge $(1/2,0)$. Therefore, the
excitations of type $\tilde{j}_{0,1}$ and $\tilde{j}_{0,2}$ are
descendant states of the highest weight states in the $[1,0]$ and
$[1,1]$ representation. From now on excitations of type
$\{j_{0,1},\tilde{j}_{0,1}\}$ are labeled as $[1,0]$-solitons while
excitations of type $\{j_{0,2},\tilde{j}_{0,2}\}$ are labeled as
$[1,1]$-solitons. The masses and ${SO}(5)$ charges of the auxiliary
modes vanish, i.e. \
${\rho}^{(m)}_{0,j_2}(\lambda) = 0 =
{\epsilon}^{(m)}_{0,j_2}(\lambda)$.

The energy density of a macro-state with densities given by (\ref{so5_densities2}) is
    \begin{equation}
        \label{so5_energy3}
		\Delta\mathcal{E}=\sum_{m=1}^{2}\sum_{j}\int_{-\infty}^{\infty}\text{d}\lambda\,{\epsilon}^{(m)}_{0,j}(\lambda)\rho^{(m)}_{j}(\lambda)\,.
    \end{equation}
Furthermore, it is convenient to define the masses $M^{(m)}_k$ of the different solitons as
\begin{equation}
 	\begin{aligned}\nonumber
 		M^{(m)}_{k} &\equiv 
 		\begin{cases}
 			2\sqrt{3}M_{0} & \text{if }m=1,\,k\in\{j_{0,1},\tilde{j}_{0,1}\}\,,\\
 			2M_{0} &  \text{if }m=2,\,k\in\{j_{0,2},\tilde{j}_{0,2}\}
 		\end{cases}\,.\nonumber
 	\end{aligned}
 \end{equation}
	

\subsection{Low-temperature thermodynamics}
\label{sec:SO5_Thermodynamics}
To derive the physical properties of the different quasi-particles
appearing in the Bethe ansatz solution of the model
(\ref{so5_NfModel}) its low-temperature thermodynamics is studied. The
equilibrium state at finite temperature is obtained by minimizing the
free energy, $F/N=\mathcal{E}-T\mathcal{S}$, with the combinatorial
entropy \cite{YaYa69}
\begin{equation}
  \label{so5_entropy}
  \mathcal{S}=\sum_{j\geq 1}\int_{-\infty}^{+\infty}\text{d}\lambda\,
  \left[(\rho_j+\rho^h_j)\ln(\rho_j+\rho^h_j)
        -\rho_j\ln\rho_j-\rho^h_j\ln\rho^h_j\right]\,.
\end{equation}
The resulting thermodynamic Bethe ansatz (TBA) equations read
\begin{align}
  \label{so5_dressedeinteq}
  &T\ln(1+e^{\epsilon^{(m)}_k/T})=
    \epsilon^{(m)}_{0,k}(\lambda) +
    \sum_{l=1}^{2}\sum_{j\geq 1}
    B^{(l,m)}_{jk}\ast T\ln(1+e^{-\epsilon^{(l)}_j/T})\,, 
\end{align}
where the dressed energies $\epsilon^{(m)}_j(\lambda)$ have been
introduced through
$e^{-\epsilon^{(m)}_j/T}= \rho^{(m)}_j/\rho^{h(m)}_j$. To study the
properties of free and interacting solitons it is convenient to
rewrite the integral equations of the auxiliary modes:
the auxiliary modes become independent of $\lambda$ for temperatures
small compared to the soliton gaps,
$T \ll \epsilon^{(m)}_{j_{0,m}}(0)$.
Similarly, they take constant values for finite values of $\lambda$ in
the condensed phases when $T \ll |\epsilon^{(m)}_{j_{0,m}}(0)|$.
In these cases the effective equations describing the auxiliary modes
are
\begin{align}
  \label{so5_auxequations}
  \begin{split}
    \epsilon^{(1)}_{j_{2,1}}=&\delta_{j_{2,1},N_f-1}T\ln\left(1+e^{-\epsilon^{(1)}_{j_{0,1}}/T}\right)^{\frac{1}{2}}+T\ln \left(1+e^{\epsilon^{(1)}_{j_{2,1}-1}/T}\right)^{\frac{1}{2}}\left(1+e^{\epsilon^{(1)}_{j_{2,1}+1}/T}\right)^{\frac{1}{2}}\\
    &-T\ln \left(1+e^{-\epsilon^{(2)}_{2j_{2,1}-1}/T}\right)^{\frac{1}{2}}\left(1+e^{-\epsilon^{(2)}_{2j_{2,1}}/T}\right)\left(1+e^{-\epsilon^{(2)}_{2j_{2,1}+1}/T}\right)^{\frac{1}{2}}\,,\\
    \epsilon^{(2)}_{j_{2,2}}=&\delta_{j_{2,2},2N_f-1}T\ln\left(1+e^{-\epsilon^{(2)}_{j_{0,2}}/T}\right)^{\frac{1}{2}}+T\ln \left(1+e^{\epsilon^{(2)}_{j_{2,2}-1}/T}\right)^{\frac{1}{2}}\left(1+e^{\epsilon^{(2)}_{j_{2,2}+1}/T}\right)^{\frac{1}{2}}\\
    &-T\ln \left(1+e^{-\epsilon^{(1)}_{j_{2,2}/2}/T}\right)^{\frac{1}{2}}\,,
  \end{split}
\end{align}
where $\epsilon^{(1)}_0=\epsilon^{(2)}_{0}=-\infty$ and
$\epsilon^{(1)}_{j_{2,2}/2}=\infty$ if $j_{2,2}$ is odd. In terms of
the dressed energies the free energy per particle is
\begin{equation}
  \label{so5_freeenergy}
  \begin{aligned}
    \frac{F}{\mathcal{N}}&\,\,=\,\,-T\sum_{m=1}^{2}\sum_{j \notin\{j_{2,m}\}} \int_{-\infty}^{\infty}\text{d}\lambda\, \rho^{(m)}_{0,j}(\lambda)\ln(1+e^{-\epsilon^{(m)}_{j}(\lambda)/T})\,\\
    &\stackrel{g\ll 1}{=}-\frac{T}{6}\sum_{m=1}^{2}\sum_{j \notin\{j_{2,m}\}} M^{(m)}_{j}\int_{-\infty}^{\infty}\text{d}\lambda\, \cosh(\pi\lambda/3)\ln(1+e^{-\epsilon^{(m)}_{j}(\lambda)/T})\,.
  \end{aligned}
\end{equation}

\begin{figure}[ht]%
  \centering
  \captionsetup[subfigure]{singlelinecheck=on}
  \subfloat[]{{\includegraphics[width=0.5\textwidth]{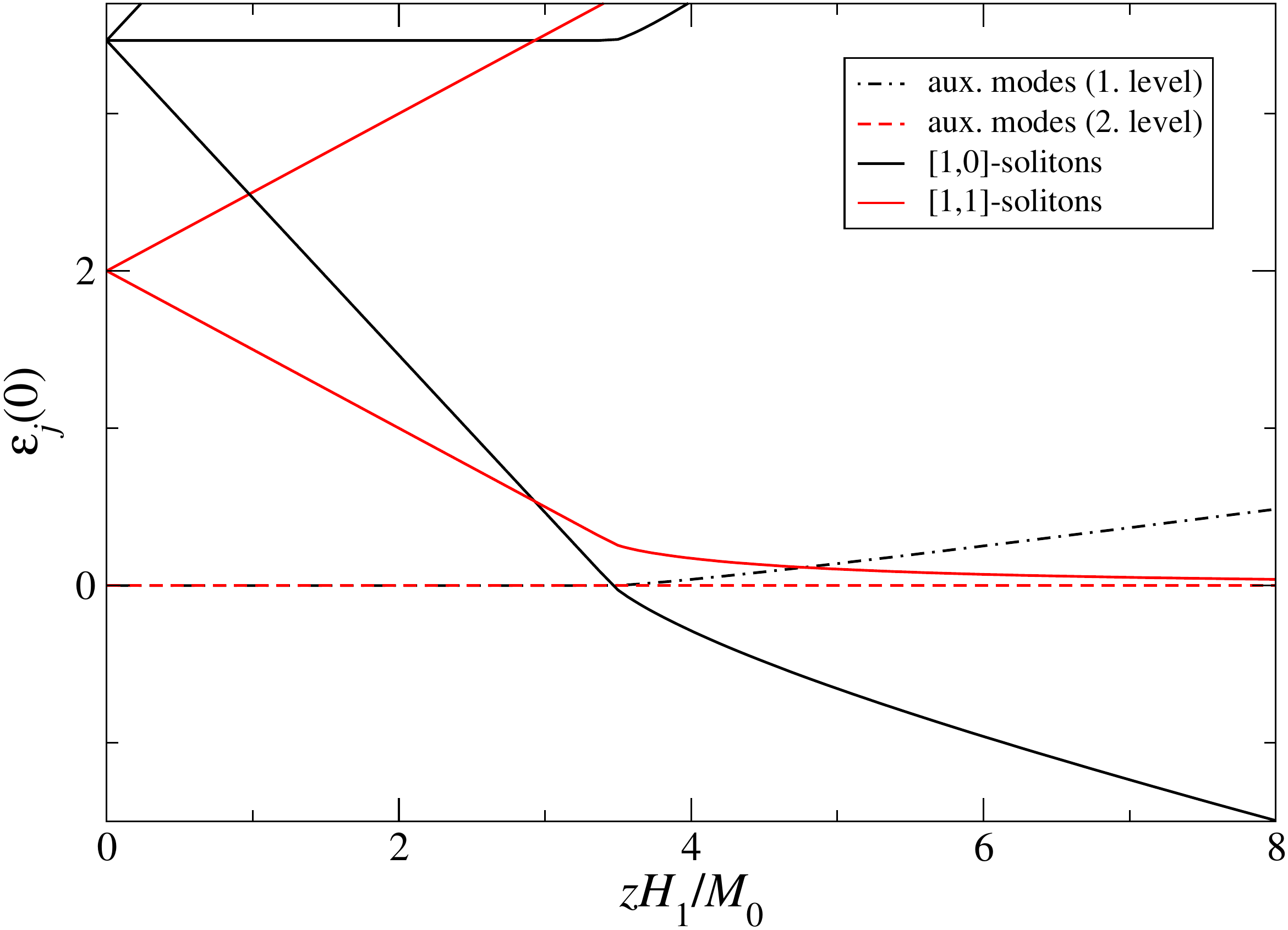} }}%
  \subfloat[]{{\includegraphics[width=0.5\textwidth]{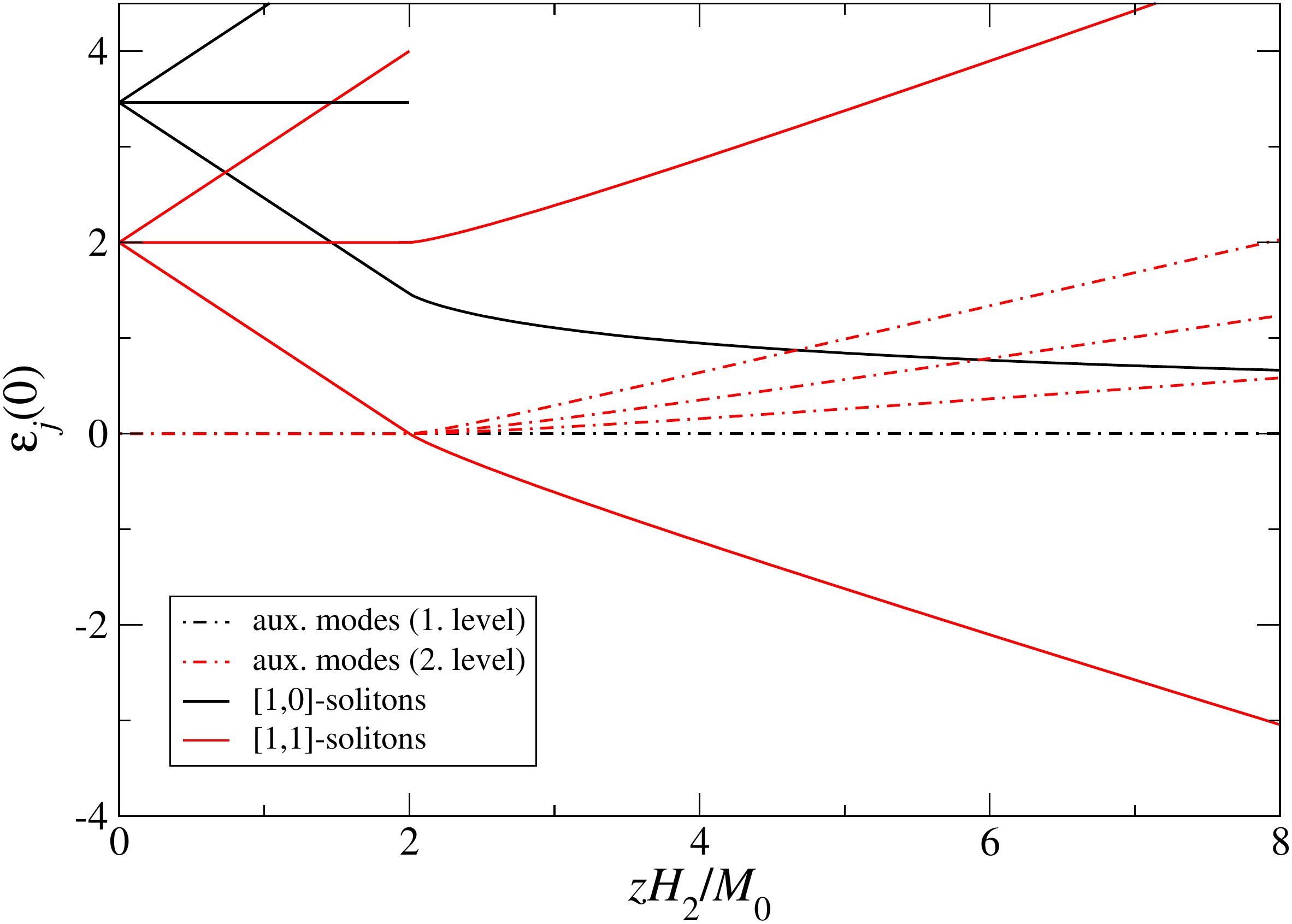} }}%

  \subfloat[]{{\includegraphics[width=0.5\textwidth]{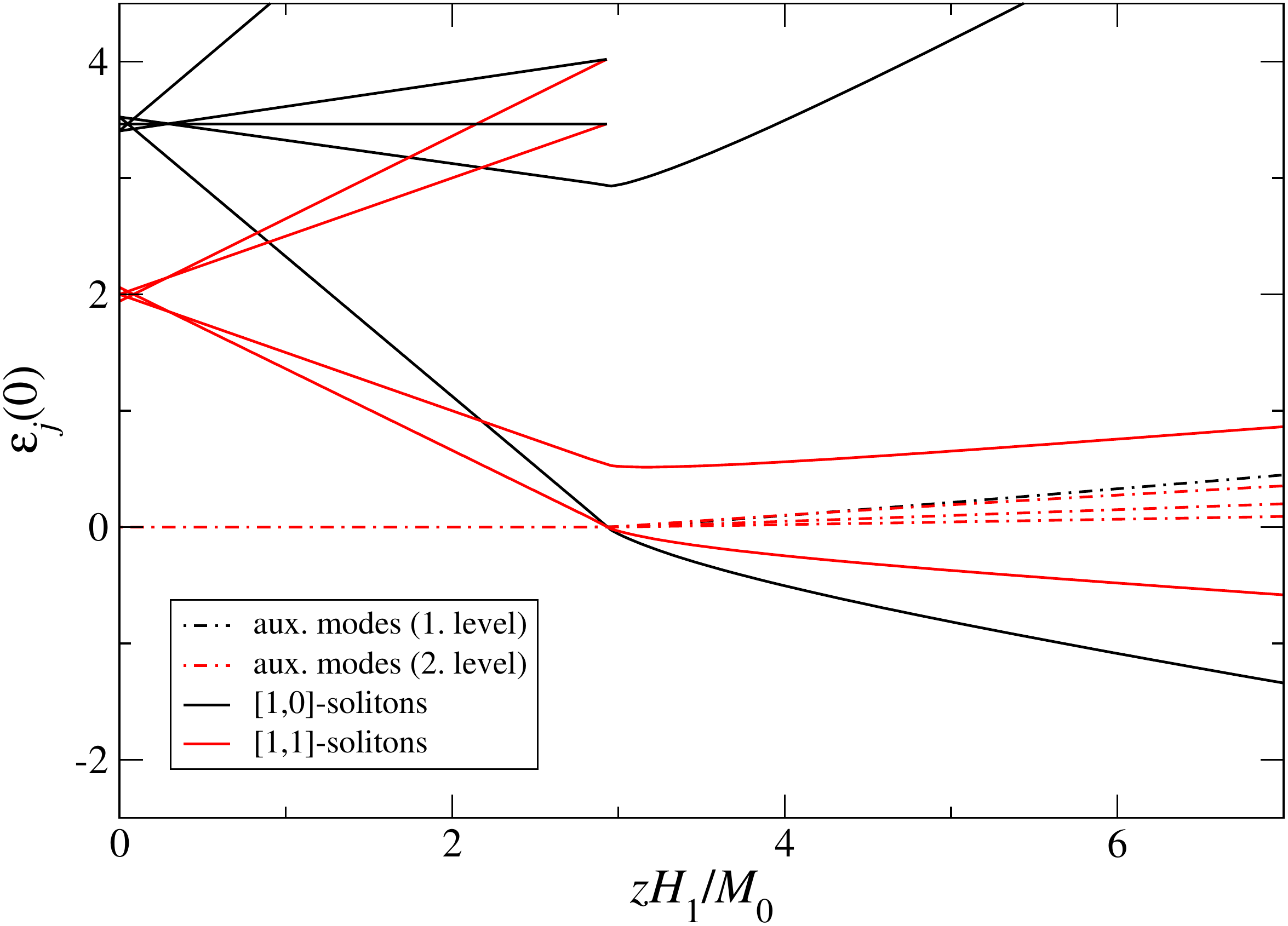} }}
  \caption{The zero temperature spectrum of elementary excitations
    (and Fermi energy of solitons in the condensed phases) 
    $\epsilon^{(m)}_j(0)$ obtained from the numerical solution of
    (\ref{so5_dressedeinteq}) for $p_0=2+1/4$: (a) for $H_2=0$ as a
    function of the field $H_1$, (b) for $H_1=0$ as function of $H_2$,
    and (c) for $zH_2/M_0=-0.06+0.21\,zH_1/M_0$ as function of $H_1$.
    Once the gap of $[1,0]$-solitons in (a) or $[1,1]$-solitons in (b)
    closes the system forms a collective state of these objects
    For sufficiently large fields both $[1,0]$- and $[1,1]$-solitons
    condense as in (c).  In these phases the degeneracy of the auxiliary
    modes is lifted.
  }%
  \label{so5_fig:spec1}%
\end{figure}

Solving the equations (\ref{so5_dressedeinteq}) the spectrum of the
model (\ref{so5_NfModel}) for a given temperature $T$ and fields
$H_1,\,H_2$ is obtained.  From the expressions (\ref{so5_energysol1})
and (\ref{so5_energysol2}) for the bare energies of the elementary
excitations the qualitative behavior of these modes at low
temperatures can be deduced, see Figure~\ref{so5_fig:phases0} for $T\to0$:
\begin{figure}[ht]%
  \centering
  \includegraphics[width=0.85\textwidth]{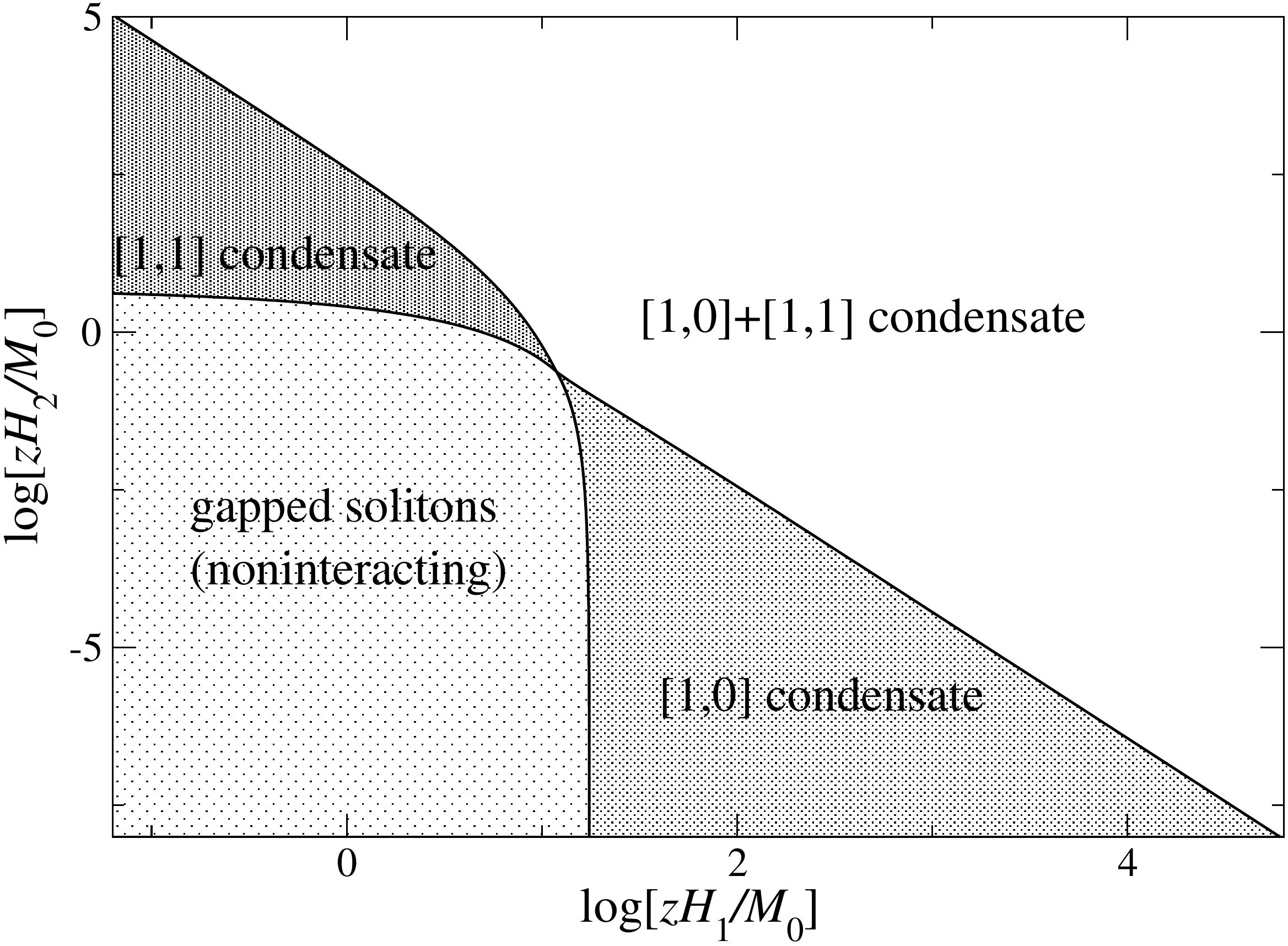}
  \caption{Zero temperature phase diagram of the model
    (\ref{so5_NfModel}: as the fields $H_{1,2}$ are increased solitons
    carrying quantum numbers of the highest weight states in the
    $[1,0]$ vector and the $[1,1]$ spinor representation of $SO(5)$
    condense (the actual location of the phase boundaries is obtained
    from the numerical solution of the TBA equations
    (\ref{so5_dressedeinteq}) for $T=0$ and $p_0=2+1/3$).}%
  \label{so5_fig:phases0}%
\end{figure}
as long as
$zH_2<\text{min}\left( 2\sqrt{3}M_0-zH_1,\, 2M_0-zH_1/2\right)$
solitons remain gapped. By increasing the field $H_1$ (above
$zH_1\geq 2\sqrt{3}M_0-zH_2$) for sufficiently small $H_2$
($zH_2<(4-2\sqrt{3})M_0$) the gap of the $[1,0]$-solitons closes and
they condense into a phase with finite density. In this collective
state the degeneracy of the auxiliary modes is lifted while the gap of
the $[1,1]$-solitons remains open until $zH_1\gg M_0$, see
Figure~\ref{so5_fig:spec1}(a) for the $T=0$ spectrum with
$H_2\equiv0$.
Similarly, for sufficiently small $H_1$ ($zH_1<4(\sqrt{3}-1)M_0$)
increasing the field $H_2$ (above $zH_2\geq 2M_0-zH_1/2$) closes the
gap of the $[1,1]$-solitons, while the gap of the $[1,0]$-solitons
remains open until $zH_2\gg M_0$, see Figure~\ref{so5_fig:spec1}(b)
for the $T=0$ spectrum with $H_1\equiv0$. In Figure~\ref{so5_fig:spec1}(c)
we display the spectrum of elementary excitations for a combination of
magnetic fields, where the gaps of $[1,0]$- and $[1,1]$-solitons close
simultaneously.

Notice that the string hypothesis (\ref{so5_string}) does not capture
all solitons of the $[1,0]$ and $[1,1]$ multiplet that may
occur. However, from the coupling of their charges to the fields the
energy gaps of all $[1,0]$- and $[1,1]$-solitons can be predicted in
the non-interacting regime, see Figures~\ref{so5_fig:spec1}.
In the following we will choose the temperatures to be sufficiently
small such that only the solitons with charges corresponding to the
highest weight states of the $[1,0]$ and $[1,1]$ multiplet, i.e.\
excitations of type $j_{0,m}$, contribute to the thermodynamics.

\subsection{Non-interacting solitons}
For fields $zH_2<\min\left(2\sqrt{3}M_0-zH_1,2M_0-zH_1/2\right)$
temperatures below the gaps of the solitons are considered, i.e.\
$T\ll
\min\left(\epsilon^{(1)}_{0,j_{0,1}}(0),\,\epsilon^{(2)}_{0,j_{0,2}}(0)\right)$. Analogously
to \cite{Tsve14a,BoFr18,BoFr18a} the nonlinear integral equations
(\ref{so5_dressedeinteq}) can be solved iteratively in this regime:
the energies $\epsilon^{(m)}_{k}$ of solitons are well described by
their first order approximation while those of the auxiliary modes can
be replaced by the asymptotic solution for
$|\lambda|\rightarrow \infty$, see Table~\ref{table:auxnumsol} for
$2\leq N_f\leq 5$.
\begin{table}[ht]
    \centering
    \bgroup
    \def\arraystretch{1.5}
    \begin{tabular}{l | l | l}
    \toprule
    $N_f$ & $\left\{\exp\left(-\epsilon^{(1)}_{j_{2,1}}/T\right)\right\}_{j_{2,1}=1}^{N_f-1}$ & $\left\{\exp\left(-\epsilon^{(2)}_{j_{2,2}}/T\right)\right\}_{j_{2,2}=1}^{2N_f-1}$\\[0.2cm]
    \hline
    $2$ & $3$ & $\frac{2}{3},\,\frac{4}{5},\,\frac{2}{3}$    \\
    $3$ & $\sqrt{3},\,\sqrt{3}$ & $\frac{1}{\sqrt{3}},\,\frac{1}{2},\,\frac{1}{3},\,\frac{1}{2},\,\frac{1}{\sqrt{3}}$   \\
    $4$ & $1.34601,\,0.91557,\,1.34601$ & $0.53679,\,0.40477,\,0.24991,\,0.27673,\,0.24991,$  \\
     &  & $0.40477,\,0.53679$ \\
     $5$ & $1.16591,\,0.66818,\,0.66818,\, 1.16591$ & $0.51415,\,0.35935,\,0.21297,\,0.20711,\,0.17157, $ \\
     & & $0.20711,\,0.21297,\,0.35935,\,0.51415$ \\
    \botrule
    \end{tabular}
    \egroup
    \caption{Asymptotic solution ($|\lambda|\rightarrow \infty$) of
      auxiliary modes
      ($\epsilon^{(m)}_{j_{2,m}}/T\equiv
      \epsilon^{(m)}_{j_{2,m}}(\lambda\rightarrow \infty)/T$) for
      $2\leq N_f\leq 5$ derived numerically from
      Eqs.~(\ref{so5_auxequations}) with
      $\epsilon^{(m)}_{j_{0,m}}/T=\infty$.}
    \label{table:auxnumsol}
\end{table}
For the other modes
    \begin{equation}\nonumber
		\epsilon^{(m)}_{j}(\lambda)=\epsilon^{(m)}_{0,j}(\lambda)-T\ln Q^{(m)}_j
\end{equation}
is obtained for $m=1,\,j\in\{j_{0,1},\tilde{j}_{0,1}\}$ and $m=2,\,j\in\{j_{0,2},\tilde{j}_{0,2}\}$ resulting in the free energy
\begin{equation}
    \label{so5_eq:Fideal}
		\frac{F}{\mathcal{N}}=-\sum_{m=1}^{2}\sum_{j \notin\{j_{2,m}\}}TQ_j^{(m)}\int\frac{\text{d}p}{2\pi}e^{-\epsilon^{(m)}_{0,j}(0)/T-p^2/2M^{(m)}_{j}T}\,,
\end{equation}
where $Q^{(m)}_{k}$ ($k\neq j_{2,m}$) depends on the asymptotic solution of the auxiliary modes
    \begin{equation}
        \label{so5_quantumdimform}
        Q^{(m)}_k=\prod_{i=1}^{2}\prod_{j_{2,i}}\left(1+e^{-\epsilon^{(i)}_{j_{2,i}}/T}\right)^{-B^{(i,m)}_{j_{2,m}k}(0)}\,.
    \end{equation}
See Table~\ref{table:so5_quantumdims} for explicit values of $Q^{(m)}_{k}$ for $2\leq N_f\leq 5$.
\begin{table}[ht]
    \centering
    \bgroup
    \def\arraystretch{1.7}
    \begin{tabular}{l | c | c}
    \toprule
    $N_f$ & $Q^{(1)}$ & $Q^{(2)}$\\[0.1cm]
    \hline
    $2$ & $2$ & $\sqrt{5}$    \\
    $3$ & $1+\sqrt{3}$ & $1+\sqrt{3}$   \\
    $4$ & $2+\sin\left({\frac{3\pi}{14}}\right)$ & $\frac{1}{2\sin\left({\frac{3\pi}{14}}\right)-1}$  \\
    $5$ & $1+\sqrt{4+2\sqrt{2}}$ & $\sqrt{2}+\sqrt{2+\sqrt{2}}$ \\
    \botrule
    \end{tabular}
    \egroup
    \caption{Quantum dimensions of the internal degrees of freedom of
      $[1,0]$-solitons ($Q^{(1)}\equiv Q^{(1)}_{j_{0,1}}$) and
      $[1,1]$-solitons ($Q^{(2)}\equiv Q^{(2)}_{j_{0,2}}$) derived
      from (\ref{so5_quantumdimform}) using the asymptotic solutions
      of the auxiliary modes (see Table~\ref{table:auxnumsol}).}
    \label{table:so5_quantumdims}
\end{table}
Following \cite{Tsve14a,BoFr18} each of the terms appearing in
Eq.~(\ref{so5_eq:Fideal}) is the free energy of an ideal gas of
particles with the corresponding mass carrying an internal degree of
freedom with possibly non-integer quantum dimension $Q^{(m)}_k$ for
the solitons. It is found that solitons of the same multiplet carry
the same quantum dimension, i.e.
$Q^{(m)}\equiv Q^{(m)}_{j_{0,m}}=Q^{(m)}_{\tilde{j}_{0,m}}$. The
densities of the solitons
\begin{equation}
        \label{so5_density}
		n^{(m)}_{j}=Q^{(m)}\sqrt{\frac{M^{(m)}_{j}T}{2\pi}}e^{-\epsilon^{(m)}_{0,j}(0)/T}\,,
\end{equation}
derived from the free energy (\ref{so5_eq:Fideal}) for $j\in\{j_{0,m},\tilde{j}_{0,m}\}$, 
can be controlled by variation of the temperature and the fields.

In order to identify the quantum dimensions $Q^{(m)}_k$ with the
quantum dimensions of ${SO}(5)_{N_f}$ anyons the topological charges
are written in terms of Young diagrams: according to \cite{FrMSbook96}
the admissible weights $\Lambda$ of the affine Lie algebra
$SO(5)_{N_f}$ have to satisfy
\begin{equation}
  \label{affinecondi}
  (\Lambda,\theta)\leq N_f\,,
\end{equation}
where $\theta$ is the highest root. In terms of the Dynkin labels
$(m_1,m_2)$, the condition (\ref{affinecondi}) results in
\begin{equation}
  \label{affinecondi2}
  m_1+m_2\leq N_f\,.
\end{equation}
Hence, $SO(5)_{N_f}$ anyons may be labeled by Dynkin labels
$(m_1,m_2)$ satisfying (\ref{affinecondi2}). Equivalently, they
can be expressed using Young diagrams. For $N_f=2$ the admissible
topological charges in terms of Young diagrams are
\begin{equation}\nonumber
  [0,0],\,[1,1],\,[1,0],\,[2,1],\,[2,0],\,[2,2]\,.
\end{equation}
The corresponding fusion rules can be found in \cite{FiFF14} using the
identification
\begin{equation}\nonumber
  \psi_1=[0,0],\,\psi_2=[1,1],\,\psi_3=[1,0],\,\psi_4=[2,1],\,\psi_5=[2,0],\,\psi=[2,2]\,.
\end{equation}
Notice that these fusion rules are consistent with the tensor product
reductions of ${SO}(5)$ irreducible representations with reasonable
modifications due to the level $N_f=2$ \footnote{An elegant graphical
  method for deriving tensor product reductions for Lie algebras with
  rank $r \leq 2$ can be found in \cite{VlRW16}.}. The quantum
dimensions extracted from the fusion rules for $N_f=2$ are given by
\begin{equation}\nonumber
  d([0,0])=d([2,2])=1,\,d([1,1])=d([2,0])=2,\,d([1,0])=d([2,1])=\sqrt{5}\,.
\end{equation}
Therefore, the appearance of the internal degrees of freedom,
$Q^{(1)}$ and $Q^{(2)}$, can be interpreted as $[1,1]$ or $[2,0]$
anyons being bound to the $[1,0]$-solitons and $[1,0]$ or $[2,1]$
anyons being bound to the $[1,1]$-solitons.

For $N_f>2$ this identification cannot be done, since the fusion rules
and quantum dimensions of ${SO}(5)_{N_f>2}$ anyons have not yet been
derived. However, following the results from the perturbed
${SU}(3)_{N_f}$ WZNW model it is conjectured that the internal degrees
of freedom, $Q^{(1)}$ and $Q^{(2)}$, coincide with the quantum
dimensions of $[1,1]$ and $[1,0]$ anyons for arbitrary $N_f\geq 2$,
respectively.

The densities of $[1,1]$ and $[1,0]$ anyons appearing in the
one-dimensional model are determined by the densities of the
corresponding solitons (\ref{so5_density}). For fields satisfying
\begin{equation}\nonumber
  zH_1>(4\sqrt{3}-4)M_0-2T\log\left(3^{1/4}\frac{Q^{(1)}}{Q^{(2)}}\right)
\end{equation}
the dominant contribution to the free energy is that of the
$[1,0]$-solitons with $[1,1]$ anyons being bound to them. In the
remaining region of non-interacting solitons the $[1,1]$-solitons with
$[1,0]$ anyons bound to them are the dominant excitations.

\subsection{Condensate of $[1,0]$-solitons}
\label{so5_sectionB}
For fields $zH_2< (4-2\sqrt{3})M_0,\, zH_1>2\sqrt{3}M_0-zH_2$ and
temperatures $T\ll zH_1+zH_2-2\sqrt{3}M_0$ the $[1,0]$-solitons (of
type $j_{0,1}$) form a condensate, while the contribution to the free
energy of the other quasi-particles can be neglected. Following
\cite{KiRe87b} we observe that the dressed energies and densities can
be related as
\begin{equation}
	\label{so5_derelations}
	\begin{aligned}
		\rho^{(m)}_j(\lambda)&=(-1)^{\delta_{j\in\{j_{2,m}\}}}\frac{1}{2\pi}\frac{\text{d}\epsilon^{(m)}_j(\lambda)}{\text{d}\lambda} f\left(\frac{\epsilon^{(m)}_j(\lambda)}{T}\right),\\
		\rho^{h(m)}_j(\lambda)&=(-1)^{\delta_{j\in\{j_{2,m}\}}}\frac{1}{2\pi}\frac{\text{d}\epsilon^{(m)}_j(\lambda)}{\text{d}\lambda} \left(1-f\left(\frac{\epsilon^{(m)}_j(\lambda)}{T}\right)\right)\,,
	\end{aligned}
\end{equation}
for $\lambda> \lambda_{\delta}$ with $\exp(\pi\lambda_{\delta}/3)\gg1$, where $f(\epsilon)= (1+e^\epsilon)^{-1}$ is the Fermi function.  Inserting this into (\ref{so5_entropy}) we get ($\phi^{(m)}_j=\epsilon^{(m)}_j/T$)
\begin{equation}
	\begin{aligned}
		\label{so5_entropy2}
	    \mathcal{S}=&-\frac{T}{\pi}\sum_{m,j}(-1)^{\delta_{j\in\{j_{2,m}\}}} \int_{\phi^{(m)}_j(\lambda_{\delta})}^{\phi^{(m)}_j(\infty)}\text{d}\phi^{(m)}_j\, \left[f(\phi^{(m)}_j)\ln f(\phi^{(m)}_j)+(1-f(\phi^{(m)}_j))\ln(1-f(\phi^{(m)}_j))\right]\,\\
		&+\sum_{m,j}\mathcal{S}^{(m)}_j(\lambda_{\delta})\, ,\\
		\mathcal{S}^{(m)}_j&(\lambda_{\delta})\equiv \int_{-\lambda_{\delta}}^{\lambda_{\delta}}\text{d}\lambda\, \left[(\rho^{(m)}_j+\rho^{h(m)}_j)\ln(\rho^{(m)}_j+\rho^{h(m)}_j)-\rho^{(m)}_j\ln\rho^{(m)}_j-\rho^{h(m)}_j\ln\rho^{h(m)}_j\right]\,.
	\end{aligned}
\end{equation}
The integrals over $\phi^{(m)}_j$ can be performed giving
	\begin{align}
		\label{so5_entropyassymp}
		&\mathcal{S}=\sum_{m,j}\mathcal{S}^{(m)}_j(\lambda_{\delta})-\frac{2T}{\pi}\sum_{m,j}(-1)^{\delta_{j\in\{j_{2,m}\}}}[L(f(\phi^{(m)}_j(\infty))-L(f(\phi^{(m)}_j(\lambda_{\delta})))]
	\end{align}
in terms of  the Rogers dilogarithm $L(x)$
\begin{equation}\nonumber
    L(x)=-\frac{1}{2}\int_{0}^{x}\text{d}y\, \left(\frac{\ln y}{1-y}+\frac{\ln (1-y)}{y}\right)\,.
\end{equation}
For large fields $zH_1\gg 2\sqrt{3}M_0-zH_2$ we have
$\log((zH_1+zH_2)/2\sqrt{3}M_0)>\lambda_{\delta}\gg1$.  Using
Eqs.~(\ref{so5_dressedeinteq}) and (\ref{so5_auxequations}) this implies
\begin{equation}
  \label{so5_ccondition}
  \begin{aligned}
    &f(\phi^{(m)}_{j_{0,m}}(\lambda_{\delta}))= \begin{cases}
      1 & \text{~for~}m=1\\
      0 & \text{~for~}m=2
    \end{cases}\,,
    &
    f(\phi^{(m)}_{j_{0,m}}(\infty))=0\,,\\
    &f(\phi^{(m)}_{\tilde{j}_{0,m}}(\lambda_{\delta}))=0\,,
    & 
    f(\phi^{(m)}_{\tilde{j}_{0,m}}(\infty))=0\,,\\
    &f(\phi^{(m)}_{j_{2,m}}(\lambda_{\delta}))=
    \begin{cases}
      0 & \text{~for~}m=1\\
      \left(\frac{\sin\left(\frac{\pi}{2N_f+2}\right)}{\sin\left(\frac{\pi(j_{2,2}+1)}{2N_f+2}\right)}\right)^2 & \text{~for~}m=2
    \end{cases}\,.
  \end{aligned}
\end{equation}
For the remaining term, $f(\phi^{(m)}_{j_{2,m}}(\infty))$, an
analytical expression is not known. However, it can be computed
numerically using the results for the asymptotic behavior of the
auxiliary modes from Table~\ref{table:auxnumsol}. From
(\ref{so5_ccondition}) one can further conclude that the densities for
$|\lambda|<\lambda_{\delta}$ are given by
\begin{equation}\nonumber
  \rho^{h(1)}_{j_{0,1}}(\lambda)=
  \rho^{(2)}_{j_{0,2}}(\lambda)=
  \rho^{(m)}_{\tilde{j}_{0,m}}(\lambda)=
  \rho^{(1)}_{j_{2,1}}=0, \quad
  \rho^{(2)}_{j_{2,2}}(\lambda)=e^{-\epsilon^{(2)}_{j_{2,2}}/T}\,
     \rho^{h(2)}_{j_{2,2}}(\lambda)\,,
\end{equation}
where $e^{-\epsilon^{(2)}_{j_{2,2}}/T}=\text{const.}$ for
$|\lambda|<\lambda_{\delta}$. Since the integral equations
(\ref{so5_densities2}) for $\rho^{h(2)}_{j_{2,2}}$ simplify in this
regime to
\begin{equation}\nonumber
  \rho^{h(2)}_{j_{2,2}}=
  -\sum_{k_{2,2}}B^{(2,2)}_{j_{2,2}k_{2,2}}
    \ast e^{-\epsilon^{(2)}_{k_{2,2}}/T}\rho^{h(2)}_{k_{2,2}}
  \qquad \text{for }|\lambda|<\lambda_{\delta},
\end{equation}
one can conclude that $\rho^{h(2)}_{j_{2,2}}\rightarrow 0$,
$\rho^{(2)}_{j_{2,2}}\rightarrow 0$ such that
$\rho^{(2)}_{j_{2,2}}/\rho^{h(2)}_{j_{2,2}}=e^{-\epsilon^{(2)}_{j_{2,2}}/T}=\text{const.}$
Consequently, $\mathcal{S}^{(m)}_j(\lambda_{\delta})=0$ for all $j,m$
is obtained. Using the Rogers dilogarithm identity
\begin{equation}
\label{dilogidentity}
  \sum_{k=2}^{n-2} L\left(\frac{\sin^2(\pi/n)}{\sin^2(\pi k/n)}\right) = \frac{\pi^2}{6}\,\frac{2(n-3)}{n}
\end{equation}
it is found that
\begin{equation}\nonumber
  \sum_{j_{2,2}}L\left(f(\phi^{(2)}_{j_{2,2}}(\lambda_{\delta})) \right)=\frac{\pi^2}{6}\left(\frac{3N_f}{N_f+1}-1 \right)\,.
\end{equation}
In general Rogers dilogarithm identities giving the relationship
between Lie algebras and central charges of parafermion conformal
field theories have only been proven for the simply laced case
\cite{Nakanishi11}. However, for the non-simply laced Lie algebra
${SO}(5)$ similar relations can be verified numerically
\begin{equation}
  \label{so5_RogerF}
  \sum_{m=1}^{2}\sum_{j_{2,m}}L\left(f(\phi^{(m)}_{j_{2,m}}(\infty)) \right)=\frac{\pi^2}{6}\left(\frac{10N_f}{N_f+3}-2 \right)\,.
\end{equation}
Hence, we obtain the following low-temperature behavior of the entropy
\begin{equation}
  \label{so5_Entropy_CFT1}
  \mathcal{S}=\frac{\pi}{3}\left(\frac{10N_f}{N_f+3}-\frac{3N_f}{N_f+1}\right)T\,
\end{equation}
which is consistent with a conformal field theory describing the collective modes given by the coset ${SO}(5)_{N_f}/{SO}(3)_{N_f}$ with central charge
\begin{equation}\nonumber
  c=\frac{10N_f}{N_f+3}-\frac{3N_f}{N_f+1}\,.
\end{equation}
Using the conformal embedding
\begin{equation}
  \label{so5_embedding1}
  \frac{{SO}(5)_{N_f}}{{SO}(3)_{N_f}}=U(1)+\frac{Z_{SO(5)_{N_f}}}{Z_{SO(3)_{N_f}}}\,
\end{equation}
where $Z_G$ denotes generalized parafermions given as the quotient
$G/U(1)^{\text{rank}(G)}$ involving the group $G$ \cite{Gepner87},
the collective modes can equivalently be described by a product of
a free $U(1)$ boson and a parafermion coset
$Z_{SO(5)_{N_f}}/Z_{SO(3)_{N_f}}$ contributing $c=1$ and
\begin{equation}\nonumber
  c=\frac{8N_f-6}{N_f+3}-\frac{2N_f-1}{N_f+1}\,,
\end{equation}
respectively.
Notice that the central charge of the coset $Z_{SO(5)_{2}}/Z_{SO(3)_{2}}$ is $c=1$, which is consistent with the results for interacting chains of $[1,1]$ $SO(5)_{2}$ anyons \cite{FiFF14}.

\begin{figure}[ht]
    \centering
    \captionsetup[subfigure]{singlelinecheck=on}
    \subfloat[]{{\includegraphics[width=7cm]{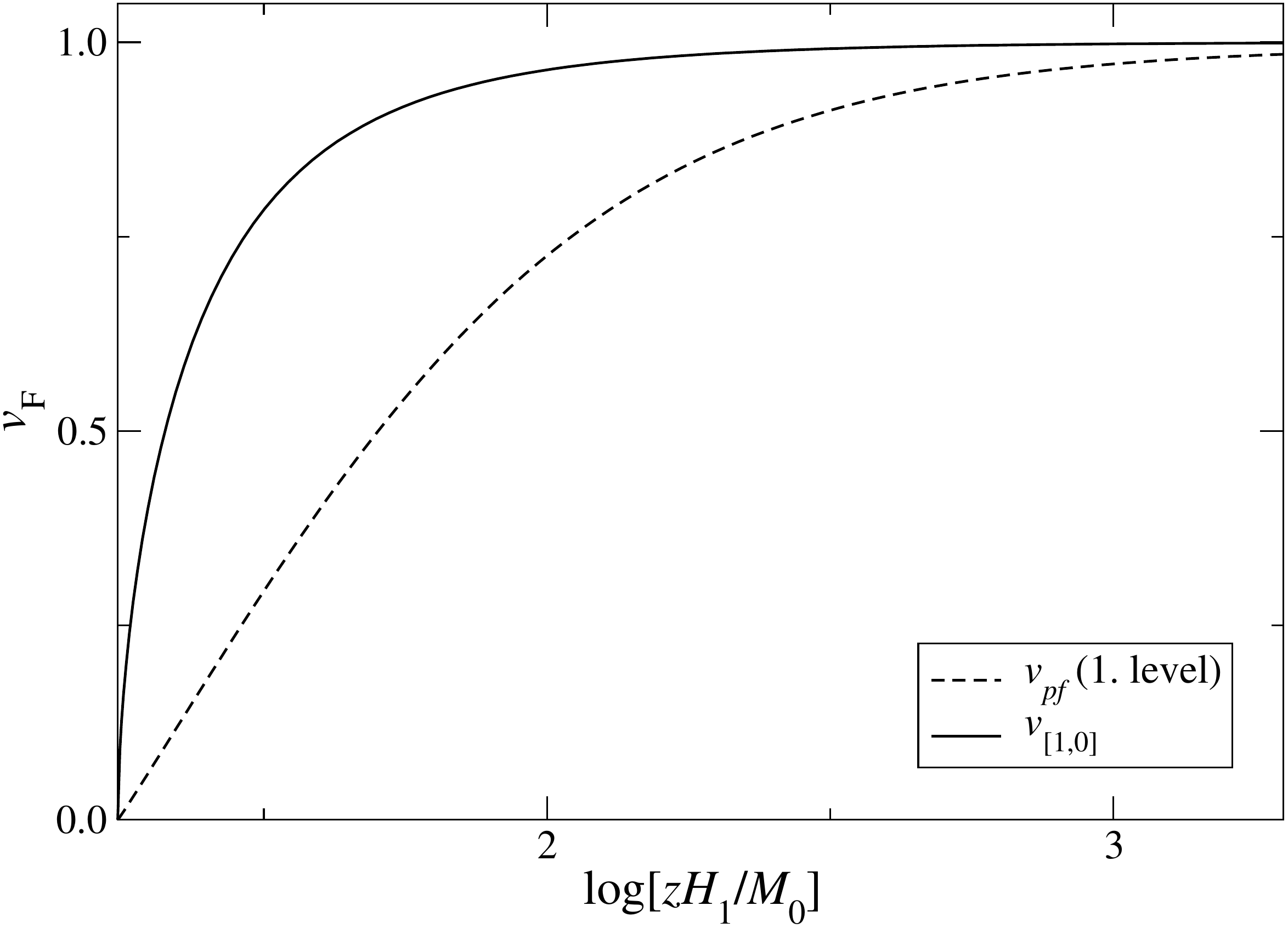} }}%
    \qquad
    \subfloat[]{{\includegraphics[width=7cm]{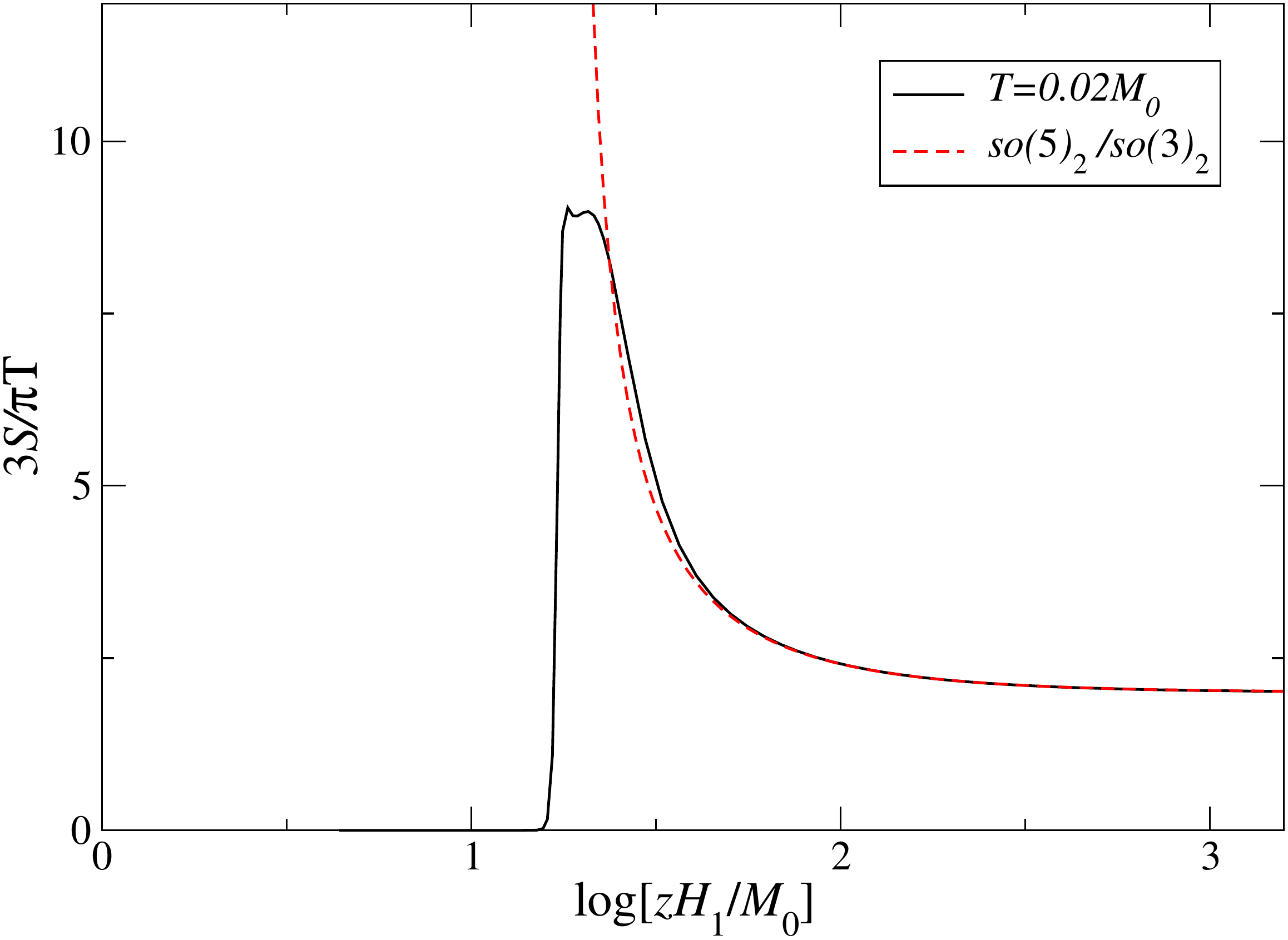} }}%
    \caption{(a) Fermi velocities of the $[1,0]$-solitons and first
      level parafermion modes as a function of the field $zH_1/M_0$
      for $p_0=2+1/3$, $H_2\equiv 0$ at zero temperature.  For large
      field, $H_1>H_{1,\delta}$, both Fermi velocities approach $1$
      leading to the asymptotic result for the low-temperature entropy
      (\ref{so5_Entropy_CFT1}). (b) Entropy obtained from numerical
      solution of the TBA equations (\ref{so5_dressedeinteq}) for
      $p_0=2+1/3$ and $H_2\equiv 0$ as a function of the field
      $zH_1/M_0$ for $T=0.02\,M_0$. For fields large compared to the
      $[1,0]$-soliton mass, $zH_1\gg 2\sqrt{3}\,M_0$, the entropy
      approaches the expected analytical value
      (\ref{so5_Entropy_CFT1}) for a field theory with a free bosonic
      sector and a $Z_{SO(5)_{N_f}}/Z_{SO(3)_{N_f}}$ parafermion
      sector propagating with velocities $v_{[1,0]}$ and
      $v^{(1)}_{pf}$, respectively (full red line).  For magnetic
      fields $zH_1<2\sqrt{3}\,M_0$ and temperature
      $T\ll 2\sqrt{3}\,M_0$ the entropy is that of a dilute gas of
      non-interacting quasi-particles with degenerate internal degree
      of freedom due to the anyons.}%
    \label{so5_fig:entropy2}%
\end{figure}

Following \cite{BoFr18} the entropy
$\mathcal{S} = -\frac{\text{d}}{\text{d}T}\frac{F}{\mathcal{N}}$ is
computed numerically to study the transition from free anyons to a
condensate of anyons.  In the region $2\sqrt{3}M_0-zH_2\lesssim zH_1$
the entropy deviates from the asymptotic expression
(\ref{so5_Entropy_CFT1}): in this range of $H_1$ the auxiliary modes
of the first level propagate with a velocity (independent of
$j_{2,1}$) differing from that of the $[1,0]$-solitons, $v_{[1,0]}$,
namely
    \begin{equation}\nonumber
    v_{[1,0]}=\left.\frac{\partial_\lambda \epsilon^{(1)}_{j_{0,1}}(\lambda)}{ 2\pi\rho^{(1)}_{j_{0,1}}(\lambda)}\right|_{\Lambda_1},\quad v^{(1)}_{pf} = -\left.\frac{ \partial_\lambda \epsilon^{(1)}_{j_{2,1}}(\lambda) }{2\pi\rho^{h(1)}_{j_{2,1}}(\lambda)}\right|_{\lambda\to\infty}\,,
\end{equation}
where $\Lambda_1$ denotes the Fermi point of $[1,0]$-solitons defined
by $\epsilon^{(1)}_{j_{0,1}}(\pm\Lambda_1)=0$. Also notice that Fermi
velocities of the second level do not exist in this regime. As a
consequence the bosonic (spinon) and parafermionic degrees of freedom
in the first level separate and the low-temperature entropy is
\begin{equation}
\label{so5_Ent_inter1}
    \mathcal{S} = \frac{\pi}{3} \left( \frac{1}{v_{[1,0]}} + \frac{1}{v^{(1)}_{pf}}\left(\frac{8N_f-6}{N_f+3}-\frac{2N_f-1}{N_f+1}\right)
    \right) T\,.
\end{equation}
This behavior can be explained by the conformal embedding
(\ref{so5_embedding1}).  Note that both Fermi velocities depend on the
field $H_1$ and approach $1$ as $H_1\gtrsim H_{1,\delta}$ such that
$\Lambda_1(H_{1,\delta})>\lambda_{\delta}$, see
Figure~\ref{so5_fig:entropy2} (a), giving the entropy
(\ref{so5_Entropy_CFT1}) of the coset $SO(5)_{N_f}/SO(3)_{N_f}$. In
Figure~\ref{so5_fig:entropy2} the computed entropy is shown for
$T=0.02\,M_0$ as a function of the field $H_1$ together with the
$T\to0$ behavior (\ref{so5_Ent_inter1}) expected from conformal field
theory.\footnote{Actually, this behavior can only be seen for
  temperatures $T<0.02\,M_0$, which was not accessible by available
  numerical methods. To overcome this problem the entropy for
  $T=0.02\,M_0$ was computed, while already neglecting the
  contribution of $\epsilon^{(2)}_{j_{0,2}}$ in the integral equations
  (\ref{so5_dressedeinteq}).}

\subsection{Condensate of $[1,1]$-solitons}
\label{so5_sectionC}
For fields $zH_1< (4\sqrt{3}-4)M_0$, $zH_2>2M_0-zH_1/2$ and
temperatures $T\ll zH_1/2+zH_2-2M_0$ the $[1,1]$-solitons (of type
$j_{0,2}$) form a condensate, while the contribution to the free
energy of the other quasi-particles can be neglected. For large fields
$zH_2\gg 2M_0-zH_1/2$
such that $\log((zH_1/2+zH_2)/2M_0))>\lambda_{\delta}\gg
1$, Eq.~(\ref{so5_dressedeinteq}) implies
\begin{equation}
  \begin{aligned}
    \label{so5_ccondition2}
    &f(\phi^{(m)}_{j_{0,m}}(\lambda_{\delta}))= \begin{cases}
      0 & \text{~for~} m=1\\
      1 & \text{~for~} m=2
    \end{cases},
    & f(\phi^{(m)}_{j_{0,m}}(\infty))=0\,,\\
    &f(\phi^{(m)}_{\tilde{j}_{0,m}}(\lambda_{\delta}))=0\,,
    &f(\phi^{(m)}_{\tilde{j}_{0,m}}(\infty))=0\,,\\
    &f(\phi^{(m)}_{j_{2,m}}(\lambda_{\delta}))=
    \begin{cases}
      \left(\frac{\sin\left(\frac{\pi}{N_f+2}\right)}{\sin\left(\frac{\pi(j_{2,1}+1)}{N_f+2}\right)}\right)^2 &  \text{~for~} m=1\\
      0 &  \text{~for~} m=2
    \end{cases}\,
  \end{aligned}
\end{equation}
together with the numerical expressions for
$f(\phi^{(m)}_{j_{2,m}}(\infty))$ obtained from the asymptotic
behavior of the auxiliary modes shown in
Table~\ref{table:auxnumsol}. The densities for
$|\lambda|<\lambda_{\delta}$ following from (\ref{so5_ccondition2})
are
\begin{equation}\nonumber
		\rho^{h(2)}_{j_{0,2}}(\lambda)=\rho^{(1)}_{j_{0,1}}(\lambda)=\rho^{(m)}_{\tilde{j}_{0,m}}(\lambda)=\rho^{(2)}_{j_{2,2}}=0, \quad \rho^{(1)}_{j_{2,1}}(\lambda)=e^{-\epsilon^{(1)}_{j_{2,1}}/T}\rho^{h(1)}_{j_{2,1}}(\lambda)\,,
\end{equation}
where $e^{-\epsilon^{(1)}_{j_{2,1}}/T}=\text{const.}$ for $|\lambda|<\lambda_{\delta}$. Since the integral equations (\ref{so5_densities2}) for $\rho^{h(1)}_{j_{2,1}}$ simplify in this regime to
    \begin{equation}\nonumber
        \rho^{h(1)}_{j_{2,1}}=-\sum_{k_{2,1}}B^{(1,1)}_{j_{2,1}k_{2,1}}\ast e^{-\epsilon^{(1)}_{k_{2,1}}/T}\rho^{h(1)}_{k_{2,1}}\qquad \text{for }|\lambda|<\lambda_{\delta},
    \end{equation}
one can conclude that $\rho^{h(1)}_{j_{2,1}}\rightarrow 0,$ $\rho^{(1)}_{j_{2,1}}\rightarrow 0$ such that $\rho^{(1)}_{j_{2,1}}/\rho^{h(1)}_{j_{2,1}}=e^{-\epsilon^{(1)}_{j_{2,1}}/T}=\text{const.}$ Consequently, $\mathcal{S}^{(m)}_j(\lambda_{\delta})=0$ for all $j,m$ is obtained. Using 
the Rogers dilogarithm identity (\ref{dilogidentity}) the relation for $Z_{SU(2)_{N_f}}$ parafermions is found:
    \begin{equation}\nonumber
        \sum_{j_{2,1}}L\left(f(\phi^{(1)}_{j_{2,1}}(\lambda_{\delta})) \right)=\frac{\pi^2}{6}\left(\frac{3N_f}{N_f+2}-1 \right)\,
    \end{equation}
Hence, the following low-temperature behavior of the entropy is obtained using (\ref{so5_RogerF})
\begin{equation}
    \label{so5_Entropy_CFT2}
    \mathcal{S}=\frac{\pi}{3}\left(\frac{10N_f}{N_f+3}-\frac{3N_f}{N_f+2}\right)T,
\end{equation}
which is consistent with a conformal field theory describing the collective modes given by the coset $SO(5)_{N_f}/SU(2)_{N_f}$ with central charge
    \begin{equation}\nonumber
        c=\frac{10N_f}{N_f+3}-\frac{3N_f}{N_f+2}.
    \end{equation}
Using the conformal embedding     
    \begin{equation}
        \nonumber
        \frac{SO(5)_{N_f}}{SU(2)_{N_f}}=U(1)+
        \frac{Z_{SO(5)_{N_f}}}{Z_{SU(2)_{N_f}}},
    \end{equation}
where $Z_{SU(N)_{N_f}}=SU(N)_{N_f}/U(1)^N$ denotes generalized $SU(N)_{N_f}$ parafermions \cite{Gepner87}, the collective modes can equivalently be described by a product of a free $U(1)$ boson and a parafermion coset $Z_{SO(5)_{N_f}}/Z_{SU(2)_{N_f}}$ contributing $c=1$ and
    \begin{equation}\nonumber
        c=\frac{8N_f-6}{N_f+3}-\frac{2(N_f-1)}{N_f+2}.
    \end{equation}
Notice that for $N_f=2$ the central charge of the coset $Z_{SO(5)_{N_f}}/Z_{SU(2)_{N_f}}$ is $c=3/2$, which is consistent with the results for interacting chains of $[1,0]$ $SO(5)_{N_f}$ anyons \cite{FiFF18}.

\begin{figure}[ht]
    \centering
    \captionsetup[subfigure]{singlelinecheck=on}
    \subfloat[]{{\includegraphics[width=7cm]{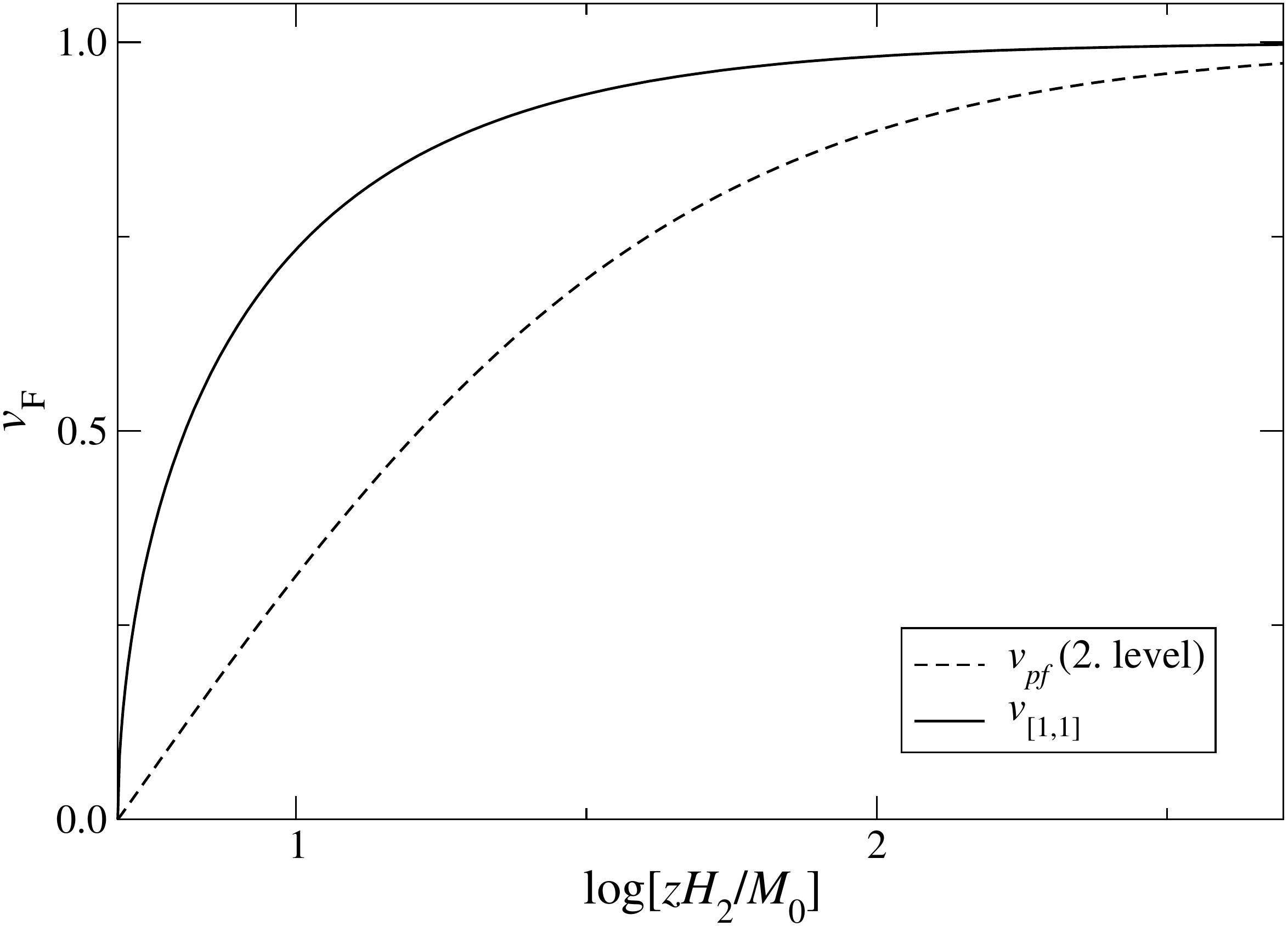} }}%
    \qquad
    \subfloat[]{{\includegraphics[width=7cm]{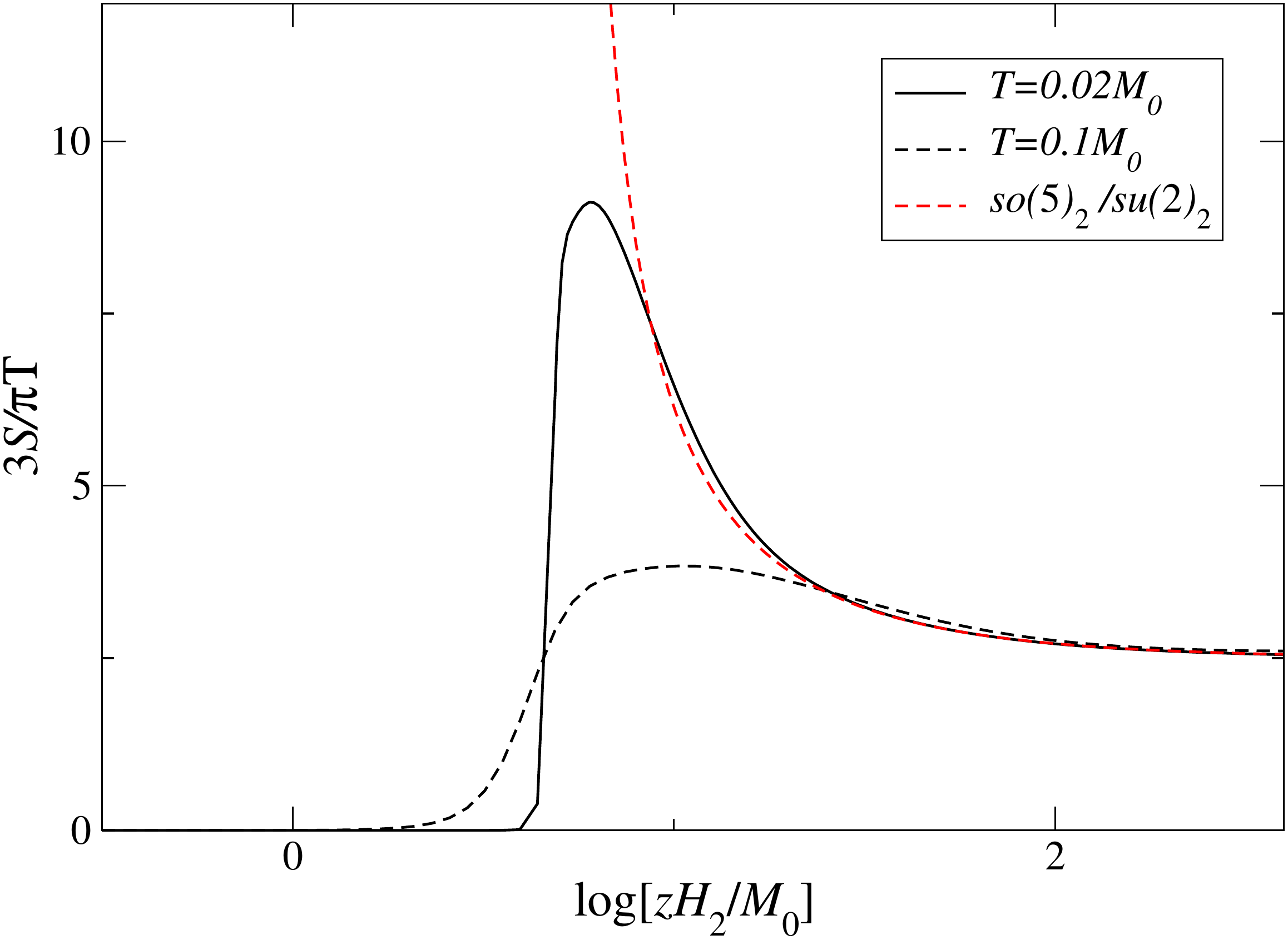} }}%
    \caption{(a) Fermi velocities of the $[1,1]$-solitons and second
      level parafermion modes as a function of the field $zH_2/M_0$
      for $p_0=2+1/3$, $H_1\equiv 0$ at zero temperature.  For large
      fields, $H_2>H_{2,\delta}$, both Fermi velocities approach $1$
      leading to the asymptotic result for the low-temperature entropy
      (\ref{so5_Entropy_CFT2}). (b) Entropy obtained from numerical
      solution of the TBA equations (\ref{so5_dressedeinteq}) for
      $p_0=2+1/3$ and $H_1\equiv 0$ as a function of the field
      $zH_2/M_{0}$ for $T=0.02M_0$. For fields large compared to the
      $[1,1]$-soliton mass, $zH_2\gg 2M_0$, the entropy approaches the
      expected analytical value (\ref{so5_Entropy_CFT2}) for a field
      theory with a free bosonic sector and a
      $Z_{SO(5)_{N_f}}/Z_{SU(2)_{N_f}}$ parafermion sector propagating
      with velocities $v_{[1,1]}$ and $v^{(2)}_{pf}$, respectively
      (full red line).  For magnetic fields $zH_2<2M_0$ and
      temperature $T\ll 2M_0$ the entropy is that of a dilute gas of
      non-interacting quasi-particles with degenerate internal degree
      of freedom due to the anyons.}%
    \label{so5_fig:entropy1}%
\end{figure}

Analogously to the regime discussed in Section~\ref{so5_sectionB}, the entropy deviates from the asymptotic expression in the region $ 2M_0-zH_1\lesssim zH_2$, since the auxiliary modes of the second level propagate with a velocity differing from that of the $[1,1]$-solitons, $v_{[1,1]}$, namely
  \begin{equation}\nonumber
    v_{[1,1]}=\left.\frac{\partial_\lambda \epsilon^{(2)}_{j_{0,2}}(\lambda)}{ 2\pi\rho^{(2)}_{j_{0,2}}(\lambda)}\right|_{\Lambda_2},\quad v^{(2)}_{pf} = -\left.\frac{ \partial_\lambda \epsilon^{(2)}_{j_{2,2}}(\lambda) }{2\pi\rho^{h(2)}_{j_{2,2}}(\lambda)}\right|_{\lambda\to\infty}\,,
\end{equation}
where $\Lambda_2$ denotes the Fermi point of $[1,1]$-solitons defined by $\epsilon^{(2)}_{j_{0,2}}(\pm\Lambda_2)=0$. Also notice that Fermi velocities of the first level do not exist in this regime. As a consequence the bosonic (spinon) and parafermionic degrees of freedom in the first level separate and the low-temperature entropy is
\begin{equation}
\label{so5_Ent_inter2}
    \mathcal{S} = \frac{\pi}{3} \left( \frac{1}{v_{[1,1]}} + \frac{1}{v^{(2)}_{pf}}\left(\frac{8N_f-6}{N_f+3}-\frac{2(N_f-1)}{N_f+2}\right)
    \right) T\,.
\end{equation}
Figure~\ref{so5_fig:entropy1} (a) shows how both Fermi velocities depend on the field $H_2$ and approach $1$ as $H_2\geq H_{2,\delta}$ such that $\Lambda_2(H_{2,\delta})>\lambda_{\delta}$. In Figure ~\ref{so5_fig:entropy1} (b) the computed entropy is shown as a function of the field $H_2$ together with the $T\rightarrow 0$ behavior expected from conformal field theory.

\subsection{Condensate of $[1,0]$- and $[1,1]$-solitons}
\label{so5_sectionD}
For fields $H_1,\,H_2$ satisfying
$zH_2>\max(2\sqrt{3}M_0-zH_1,\,2M_0-zH_1/2)$ and temperatures
$T\ll
-\min\left(\epsilon^{(1)}_{0,j_{0,1}}(0),\,\epsilon^{(2)}_{0,j_{0,2}}(0)\right)$
the highest weight $[1,0]$- and $[1,1]$-solitons condense. From
Fig.~\ref{so5_fig:spec1}(c) one can further conclude that descendent
$[1,0]$- and $[1,1]$-solitons are negligible in this regime of
temperatures and magnetic fields.

The condensation of highest weight $[1,0]$- and $[1,1]$-solitons
results in non-zero Fermi velocities
$v_{[1,0]},\,v_{[1,1]},\, v^{(m)}_{pf}$ $(m=1,2)$ for the solitons and
the auxiliary modes. For large fields $zH_1\gg M_0,\,zH_2\gg M_{0}$
the following relations are found using (\ref{so5_dressedeinteq})
\begin{equation}
	\begin{aligned}\nonumber
		&f(\phi^{(m)}_{j_{0,m}}(\lambda_{\delta}))=1,
		\quad\qquad\qquad\, f(\phi^{(m)}_{j_{0,m}}(\infty))=0,\\
		&f(\phi^{(m)}_{\tilde{j}_{0,m}}(\lambda_{\delta}))=0, \quad\qquad\qquad\, f(\phi^{(m)}_{\tilde{j}_{0,m}}(\infty))=0,\nonumber\\
		&f(\phi^{(m)}_{j_{2,m}}(\lambda_{\delta}))=0\nonumber
	\end{aligned}
\end{equation}
and therefore
\begin{equation}\nonumber
		\rho^{h(m)}_{j_{0,m}}(\lambda)=\rho^{(m)}_{j_{1,m}}(\lambda)=\rho^{(m)}_{j_{2,m}}=0,
		\quad \mathrm{for~} |\lambda|<\lambda_{\delta}\,,
\end{equation}
giving $\mathcal{S}^{(m)}_j(\lambda_{\delta})=0$ for all $j,m$. Using the relation (\ref{so5_RogerF}) the low-temperature behavior of the entropy becomes
\begin{equation}
  \label{so5_entropy_CFT3}
  \mathcal{S}=\frac{\pi}{3}\frac{10N_f}{N_f+3}T
\end{equation}
in the phase with finite $[1,0]$- and $[1,1]$-soliton density. The
low-energy excitations near the Fermi points
$\epsilon^{(m)}_{j_{0,m}}(\pm\Lambda_m)=0$ of the soliton dispersion
propagate with velocity $v_{[1,0]}=v_{[1,1]}\rightarrow 1$ for fields
$H_m>H_{m,\delta}$ such that
$\Lambda_m(H_{m,\delta})>\lambda_{\delta}$.  Hence, the conformal
field theory describing the collective low-energy modes is the $SO(5)$
WZNW model at level $N_f$ or, by conformal embedding \cite{Gepner87},
a product of two free $U(1)$ bosons and a $SO(5)$ parafermionic coset
$SO(5)_{N_f}/U(1)^2$ contributing $c=2$ and
\begin{equation}
    \label{so5_cSU3para}
	c=\frac{10N_f}{N_f+3}-2=\frac{8N_f-6}{N_f+3}\,
\end{equation}
to the central charge, respectively.

\begin{figure}[ht]
    \centering
    \captionsetup[subfigure]{singlelinecheck=on}
    \subfloat[]{{\includegraphics[width=7cm]{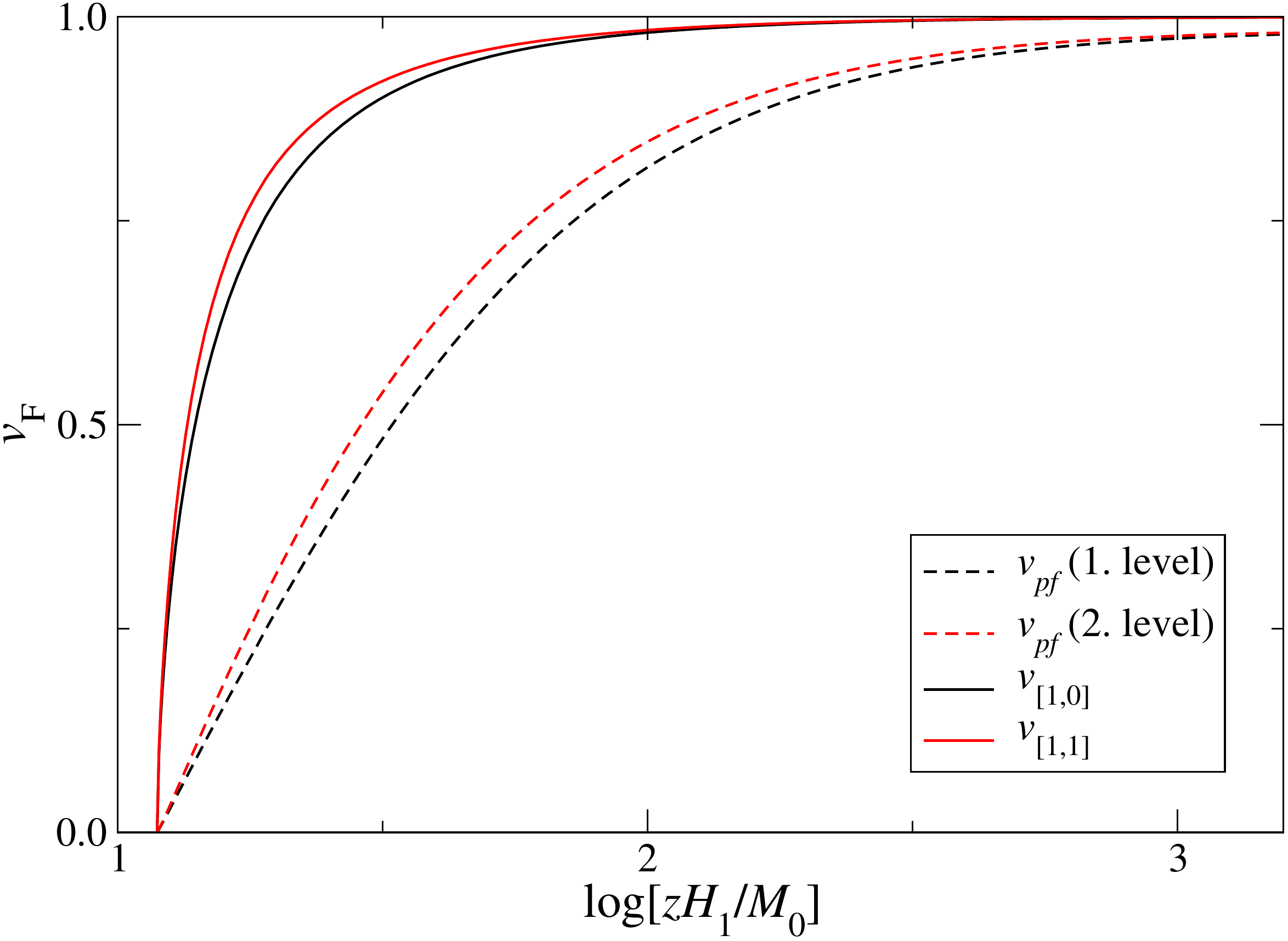} }}%
    \qquad
    \subfloat[]{{\includegraphics[width=7cm]{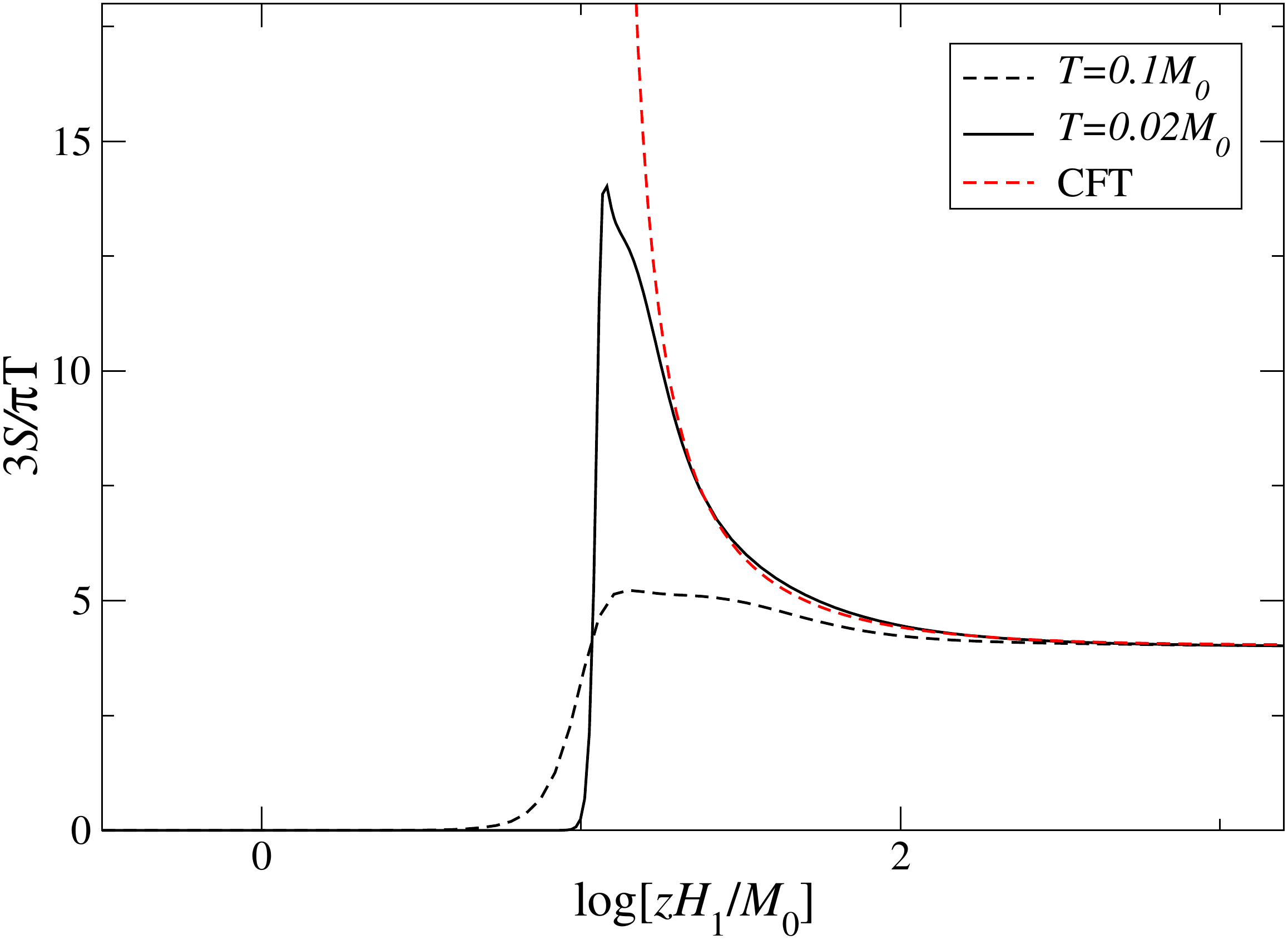} }}%
    \caption{(a) Fermi velocities as a function of the field
      $zH_1/M_0$ for $p_0=2+1/3,\,$ $zH_2=-0.06M_0+0.21zH_1$ at zero
      temperature.  For large fields, $H_1>H_{1,\delta}$, all Fermi
      velocities approach $1$ leading to the asymptotic result for the
      low-temperature entropy (\ref{so5_entropy_CFT3}). (b) Entropy
      obtained from numerical solution of the TBA equations
      (\ref{so5_dressedeinteq}) as a function of the field
      $zH_2/M_{0}$ for $p_0=2+1/3$, fixed $zH_2=-0.06\,M_0+0.21\,zH_1$
      and different temperatures. For fields large compared to the
      kink mass, $zH_1\gg M_{0}$, the entropy approaches the expected
      analytical value (\ref{so5_entropy_CFT3}) (full red line).  For
      magnetic fields $zH_1<2(M^{(1)}_{j_{0,1}}-M^{(2)}_{j_{0,2}})$
      and temperature $T\ll M_{0}$ the entropy is that of a dilute gas
      of non-interacting quasi-particles with degenerate internal
      degree of freedom due to the anyons.}%
    \label{so5_fig:entropy3}%
\end{figure}

For fields $H_1,\, H_2$ such that $v_{[1,0]}=v_{[1,1]}<1$ and $v^{(1)}_{pf}=v^{(2)}_{pf}<1$ the degeneracy between the solitons and the parafermions is lifted resulting in the low-temperature behavior of the entropy given by
\begin{equation}\nonumber
    \mathcal{S}=\frac{\pi}{3}\left(\frac{2}{v_{[1,0]}}+\frac{1}{v^{(m)}_{pf}}\frac{8N_f-6}{N_f+3} \right)T\,.
\end{equation}
Additionally, the fields can be chosen such that the remaining degeneracies are lifted, i.e. $v_{[1,0]}<v_{[1,1]}$ and $v^{(1)}_{pf}<v^{(2)}_{pf}$. In this case the entropy becomes
\begin{equation}
    \nonumber
    \mathcal{S}=\frac{\pi}{3}\left(\frac{1}{v_{[1,0]}}+\frac{1}{v^{(1)}_{pf}}\frac{2(N_f-1)}{N_f+2}+\frac{1}{v_{[1,1]}}+\frac{1}{v^{(2)}_{pf}}\left(\frac{8N_f-6}{N_f+3}-\frac{2(N_f-1)}{N_f+2}\right) \right)T\,,
\end{equation}
which is consistent with the conformal embedding
    \begin{equation}\nonumber
        SO(5)_{N_f}=U(1)+Z_{SU(2)_{N_f}}+U(1)+\frac{Z_{SO(5)_{N_f}}}{Z_{SU(2)_{N_f}}}\, ,
    \end{equation}
see Figure~\ref{so5_fig:entropy3} (a) for the Fermi velocities and Figure~\ref{so5_fig:entropy3} (b) for the entropy in this regime.

At last, for Fermi velocities $v_{[1,1]}<v_{[1,0]}$ and
$v^{(2)}_{pf}<v^{(1)}_{pf}$ the entropy results in
\begin{equation}\nonumber
    \mathcal{S}=\frac{\pi}{3}\left(\frac{1}{v_{[1,1]}}+\frac{1}{v^{(2)}_{pf}}\frac{2N_f-1}{N_f+1}+\frac{1}{v_{[1,0]}}+\frac{1}{v^{(1)}_{pf}}\left(\frac{8N_f-6}{N_f+3}-\frac{2N_f-1}{N_f+1}\right) \right)T\,,
\end{equation}
which is consistent with the conformal embedding
\begin{equation}\nonumber
    SO(5)_{N_f}=U(1)+Z_{SO(3)_{N_f}}+U(1)+\frac{Z_{SO(5)_{N_f}}}{Z_{SO(3)_{N_f}}}\,.
\end{equation}

\newpage

\section{Summary and conclusion}
\label{sec:SO5_Summary}

\begin{figure}[ht]\centering
    \includegraphics[width=0.85\textwidth]{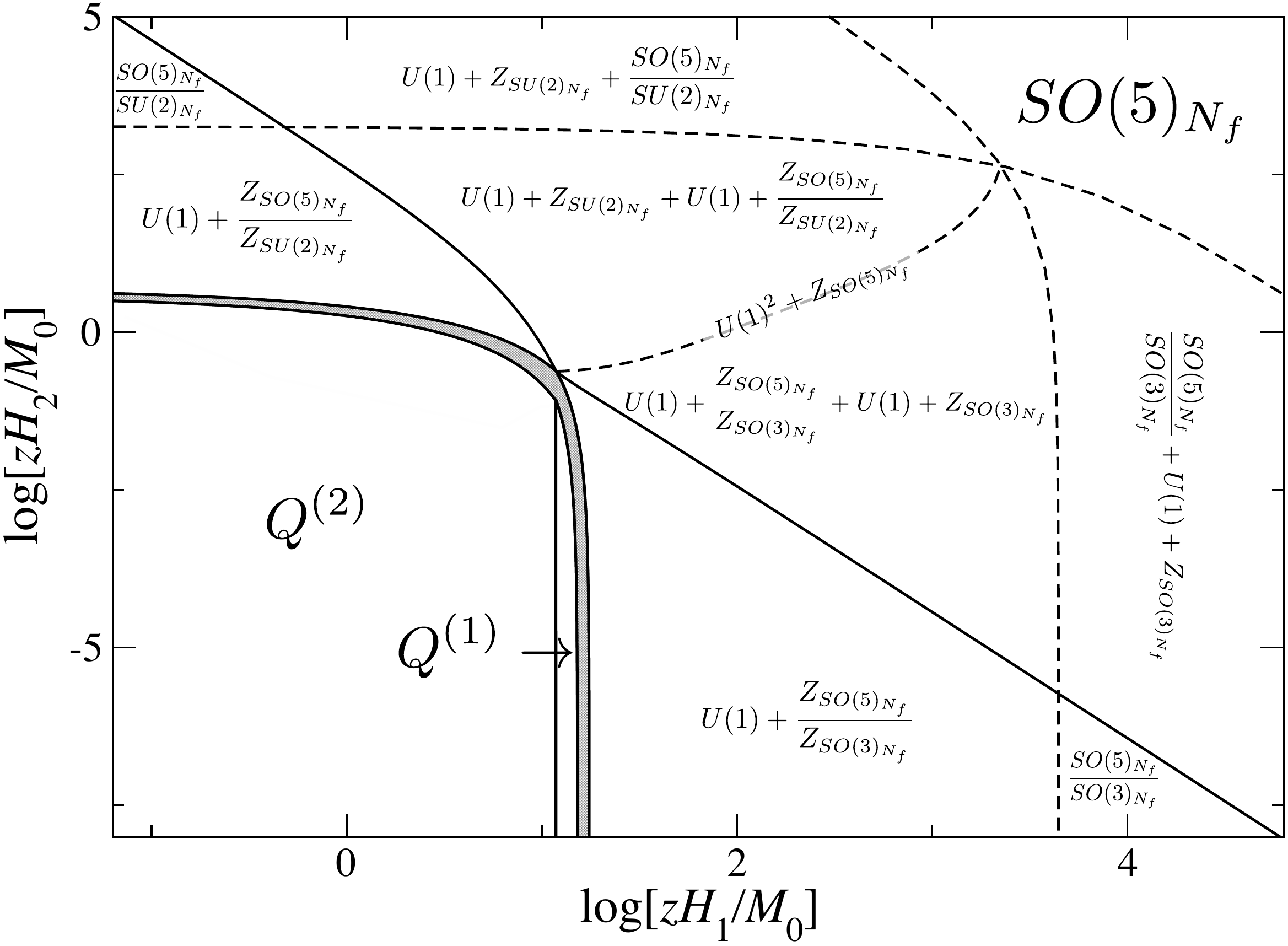}
    \caption{ Contribution of the $SO(5)_{N_f}$ anyons to the
      low-temperature properties of the model (\ref{so5_NfModel}), see
      Figure~\ref{so5_fig:phases0} for the phases of the solitonic
      quasi-partiles: using the criteria described in the main text
      the parameter regions are identified using analytical arguments
      for $T\to0$ (the actual location of the boundaries is based on
      numerical data for $p_0=2+1/3$ and $T=0.035\,M_0$). For small
      fields (regions $Q^{(1)}$, $Q^{(2)}$) a dilute gas of
      non-interacting quasi-particles with an internal anyonic
      (zero-energy) degree of freedom with quantum dimension $Q^{(1)}$
      or $Q^{(2)}$ is realized.  In the shaded region The degeneracy
      of the zero modes is lifted by the presence of thermally
      activated solitons with a small but finite density. All the
      other phases are labelled by the CFT describing the collective
      state formed by the condensed degrees of freedom.
    \label{so5_fig:phasediag}}
\end{figure}

Our findings are summarized in a phase diagram based on the numerical
analysis of the TBA equations (\ref{so5_dressedeinteq}), see
Figure~\ref{so5_fig:phasediag}. For sufficiently small fields a dilute
gas of anyons with quantum dimension $Q^{(1)}$ or $Q^{(2)}$ is
dominating the contribution to the free energy. By varying the
magnetic fields the condensation of anyons can be driven into various
collective states described by parafermionic cosets: the collective
state describing the condensation of $[1,1]$ $SO(5)_{N_f}$ anyons is
identified as the $Z_{SO(5)_{N_f}}/Z_{SO(3)_{N_f}}$ parafermion coset,
while the condensation of $[1,0]$ $SO(5)_{N_f}$ anyons results in the
$Z_{SO(5)_{N_f}}/Z_{SU(2)_{N_f}}$ parafermionic theory. Moreover, the
condensation of a mixture of $[1,0]$ and $[1,1]$ anyons is studied
resulting in the $Z_{SO(5)_{N_f}}$ parafermion theory describing the
collective state. Other theories describing the condensation of
$SO(5)_{N_f}$ anyons are based on conformal embeddings, see
Figure~\ref{so5_fig:phasediag}.

In summary we can conclude that the effective model describing the
$SO(5)$ spin excitations is the $SO(5)_{N_f}$ WZNW model with an
anisotropic current-current perturbation. In contrast to the previous
application of this approach to $SU(k)_{N_f}$ anyons in
\cite{BoFr18,BoFr18a} this was not clear from the beginning, since
corresponding non-Abelian bosonization results of free fermions with
$SO(5)$ spin and $U(N_f)$ flavour degrees of freedom are missing.

\begin{acknowledgments}
  Funding for this work has been provided by the \emph{School for
    Contacts in Nanosystems}.  Additional support by the research unit
  \emph{Correlations in Integrable Quantum Many-Body Systems}
  (FOR2316) is gratefully acknowledged.
\end{acknowledgments}

\appendix

\section{Representation theory of $SO(5)$}
\label{app:SO5}
The algebra $SO(5)$ has dimension $10$ and rank two.  In terms of the self
adjoint generators $M_{ab}=-M_{ba}$, $a,b=1,2,\ldots,5$, the
commutation relations of the algebra read
\begin{equation}
  \label{so5_comm}
  \left[M_{ab},M_{cd}\right] = -i \left(
    \delta_{bc}\,M_{ad} - \delta_{ac}\,M_{bd}
    - \delta_{bd}\,M_{ac} + \delta_{ad}\,M_{bc}\right)
\end{equation}
We choose the generators of the Cartan subalgebra to be
$H_c^1=M_{1,2}$ and $H_c^2=M_{3,4}$.
The $SO(5)$ root diagram is
\begin{center}
\begin{tikzpicture}[scale=1.5,>=stealth]
  \draw[->, thin]  (-1.8,0) -- (1.8,0);
  \coordinate [label=right:{$H_c^1$}] () at (1.8,0);
  \draw[->, thin]  (0,-1.5) -- (0,1.5);
  \coordinate [label=above:{$H_c^2$}] () at (0,1.5);
  
  \draw[->, very thick]  (0,0) -- (1,-1);
  \coordinate [label=above right:{$\alpha^1$}] () at (1,-1);
  \draw[->, very thick]  (0,0) -- (0,1);
  \coordinate [label=above right:{$\alpha^2$}] () at (0,1);
  
  \draw[->, dashed, very thick]  (0,0) -- (1,0);
  \coordinate [label=above:{$\alpha^1+\alpha^2$}] () at (1.3,0);
  \draw[->, dashed, thick]  (0,0) -- (1,1);
  \coordinate [label=above:{$\alpha^1+2\alpha^2$}] () at (1.3,1);

  \draw[->, dashed, very thick]  (0,0) -- (-1,1);
  \draw[->, dashed, very thick]  (0,0) -- (-1,0);
  \draw[->, dashed, very thick]  (0,0) -- (-1,-1);
  \draw[->, dashed, very thick]  (0,0) -- (0,-1);
\end{tikzpicture}
\end{center}
The corresponding ladder operators $E_\alpha$ used in the construction
of the Hamiltonian (\ref{so5_NfModel}) are linear combinations of the
other generators, e.g.\
$E_{(\pm1,0)}= \left(M_{1,5}\pm iM_{2,5}\right)$.
The two simple roots $\alpha^1=(1,-1)$ and $\alpha^2=(0,1)$ have
different length and the Cartan matrix is
\begin{equation}
  A = \left( \begin{array}{rr} 2 & -2 \\ -1 & 2 
             \end{array}\right)\,.
\end{equation}
The fundamental weights are
\begin{equation}
  \label{so5_fundw}
  \omega^1 = e^1\,,\quad \omega^2 = \frac12(e^1+e^2)\,.
\end{equation}
Equivalently, these representations can be labelled by their Dynkin
labels $(1,0)$ and $(0,1)$ or Young diagrams $[1,0]$ and $[1,1]$,
respectively (the diagram $[x_1,x_2]$ consists of $x_i$ nodes in the
$i$-th row).  The generators in the five dimensional vector
representation corresponding to $\omega^1$ are
\begin{equation}
  \left[M_{ab}\right]_{xy} = -i \left( \delta_{ax}\delta_{by}
    - \delta_{bx}\delta_{ay}\right)\,,
\end{equation}
while the four dimensional spinor representation $\omega^2$ is built
from tensor products of Pauli matrices
\begin{equation}
  \begin{aligned}
    M_{1,5} &= \frac12 \sigma^x\otimes \mathbf{1}\,,\quad
    &&M_{2,5}=\frac12 \sigma^y\otimes \mathbf{1}\,,\\
    M_{3,5} &= \frac12  \sigma^z\otimes\sigma^x\,,\quad
    &&M_{4,5}=\frac12  \sigma^z\otimes\sigma^y\,.
  \end{aligned}
\end{equation}
The other generators can be obtained from the commutation relations
(\ref{so5_comm}) giving, e.g., the Cartan generators
$H_c^1=\frac12 \sigma^z\otimes \mathbf{1}$ and
$H_c^2=\frac12 \mathbf{1}\otimes\sigma^z$.

\section{TBA of the perturbed $SO(5)_{N_f}$ WZNW model}
\label{so5_app:onTBA}
In order to obtain the integral equations (\ref{so5_densities2}) a root configuration consisting of $\nu^{(m)}_j$ strings of type $(n^{(m)}_j,v^{(m)}_{n_j})$ on the $m$-th level is considered and the Bethe equations (\ref{so5_betheeq}) are rewritten in terms of the real string-centers $\lambda^{(m,j)}_\alpha \equiv \lambda^{(m)n_j}_\alpha$ using (\ref{so5_string}).  In their logarithmic form they read
    \begin{equation}
        \label{so5_centereq}
        \begin{aligned}
            \sum_{\tau=\pm 1}\frac{N}{2}t_{k,N_f}(\lambda^{(1,k)}_\alpha+\tau/g)&=2\pi I^{(1,k)}_\alpha+\sum_{m=1}^2\sum_j\sum_{\beta=1}^{\nu^{(m)}_j}(-1)^{m+1}\theta^{(1,m)}_{kj}(\lambda^{(1,k)}_\alpha-\lambda^{(m,j)}_\beta)\,,\\
            0 &= 2\pi I^{(2,k)}_\alpha +\sum_{m=1}^2\sum_j\sum_{\beta=1}^{\nu^{(m)}_j}(-1)^{m}\theta^{(2,m)}_{kj}(\lambda^{(2,k)}_\alpha-\lambda^{(m,j)}_\beta)\,,
        \end{aligned}
    \end{equation}
where $I^{(m,k)}_\alpha$ are integers (or half-integers) and the functions
    \begin{equation}
        \label{so5_kernelrel1}
        \begin{aligned}
            t_{k,N_f}(\lambda)=&\sum_{l=1}^{\min(n_k,N_f)}f(\lambda,|n_k-N_f|+2l-1,v_kv_{N_f})\,,\\
        \theta^{(m,m)}_{kj}(\lambda)=&f(\lambda,|n^{(m)}_k-n^{(m)}_j|,v^{(m)}_kv^{(m)}_j)+f(\lambda,n^{(m)}_k+n^{(m)}_j,v^{(m)}_kv^{(m)}_j)\\
        &+2\sum_{\ell=1}^{\min(n^{(m)}_k,n^{(m)}_j)-1}f(\lambda,|n^{(m)}_k-n^{(m)}_j|+2\ell,v^{(m)}_kv^{(m)}_j) \qquad \text{with }m=1,2\,,\\
        \theta^{(1,2)}_{kj}(\lambda)=&\sum_{l=1}^{\min(2n^{(1)}_k,n^{(2)}_j)}f(\lambda,|n^{(2)}_j/2-n^{(1)}_k|+2l-1,v^{(1)}_kv^{(2)}_j)\,,\\ \theta^{(2,1)}_{kj}(\lambda)\equiv& \theta^{(1,2)}_{jk}(\lambda)
        \end{aligned}
    \end{equation}
were introduced with
\begin{equation*}
    f(\lambda,n,v)=
		\begin{cases}
			 2\arctan\left(\tan((\frac{1+v}{4}-\frac{n}{2p_0})\pi)\tanh(\frac{\pi\lambda}{2p_0})\right) \quad \text{if }\frac{n}{p_0}\neq \text{integer} \\
			 0 \hspace*{6.8cm} \text{if }\frac{n}{p_0}= \text{integer}
		\end{cases}\,.
\end{equation*}
In the thermodynamic limit, $N_m,\mathcal{N}\rightarrow \infty$ with
$N_m/\mathcal{N}$ fixed, the centers $\lambda^{(m,k)}_\alpha$ are
distributed continuously with densities $\rho^{(m)}_k(\lambda)$ and
hole densities $\rho^{h(m)}_k(\lambda)$. Following \cite{YaYa66a} the
densities are defined through the following integral equations
    \begin{equation}
        \label{so5_densities1}
        \begin{aligned}
            \tilde{\rho}^{(1)}_{0,k}(\lambda) &= (-1)^{r^{(1)}(k)}\rho^{h(1)}_k(\lambda)+\sum_{m=1}^2\sum_j(-1)^{m+1}A^{(1,m)}_{kj}\ast \rho^{(m)}_j(\lambda) \quad \text{with }k=1,\dots,j_{0,1}\,,\\
            0 &= (-1)^{r^{(2)}(k)}\rho^{h(2)}_k(\lambda)+\sum_{m=1}^2\sum_j(-1)^{m}A^{(2,m)}_{kj}\ast \rho^{(m)}_j(\lambda) \quad \text{with }k=1,\dots,j_{0,2}\,,
        \end{aligned}
    \end{equation}
where $a\ast b$ denotes a convolution and $r^{(m)}(j)$ is given by
    \begin{equation}\nonumber
		r^{(m)}(j_{2,m})=0\,,\qquad r^{(m)}(\tilde{j}_{0,m})=1\,, \qquad 
		    r^{(m)}(j_{0,m})=2\,.\nonumber
	\end{equation}
The bare densities $\tilde{\rho}^{(1)}_{0,j}(\lambda)$ and the kernels $A^{(m)}_{jk}(\lambda)$ of the integral equations are defined by
\begin{equation}
    \label{so5_kernelrel2}
	\begin{aligned}
		\tilde{\rho}^{(1)}_{0,j}(\lambda)&=\frac{1}{2}\left(a_{j,N_f}(\lambda+1/g)+a_{j,N_f}(\lambda-1/g)\right)\,, \quad
		a_{j,N_f}(\lambda)=\frac{1}{2\pi}\frac{d}{d\lambda}t_{j,N_f}(\lambda)\,,\\
		A^{(m,l)}_{kj}(\lambda)&=\frac{1}{2\pi}\frac{d}{d\lambda}\theta^{(m,l)}_{kj}(\lambda)+(-1)^{r^{(m)}(k)}\delta_{m,l}\delta_{jk}\delta(\lambda)\,.
	\end{aligned}
\end{equation}
Using (\ref{so5_energy}) and the solutions $\rho^{(m)}_k$ of (\ref{so5_densities1}) the energy density $\mathcal{E}=E/\mathcal{N}$ is rewritten as
    \begin{equation}
	\label{so5_energy2}
	\begin{aligned}
		\mathcal{E}&=\frac{1}{\mathcal{N}}\sum_{j}\sum_{\alpha=1}^{\nu^{(1)}_j}\left(\sum_{\tau=\pm 1}\frac{\tau}{2}t_{j,N_f}(\lambda^{(j,1)}_{\alpha}+\tau/g)+n_jH_1\right)+\frac{1}{\mathcal{N}}\sum_j\sum_{\alpha=1}^{\nu^{(2)}_j}n_jH_2-H_1-H_2\\
		&\stackrel{\mathcal{N}\rightarrow \infty}{=}\sum^2_{m=1}\sum_{j\geq 1}\int_{-\infty}^{+\infty}\text{d}\lambda\, \tilde{\epsilon}^{(m)}_{0,j}(\lambda)\rho^{(m)}_{j}(\lambda)-H_1-H_2\,,
	\end{aligned}
\end{equation}
where the bare energies
\begin{equation}\nonumber
  \tilde{\epsilon}^{(1)}_{0,j}(\lambda)=\sum_{\tau=\pm 1}\frac{\tau}{2}t_{j,N_f}(\lambda+\tau/g)+n_jH_1, \quad
  \tilde{\epsilon}^{(2)}_{0,j}(\lambda)=n_jH_2
\end{equation}
were introduced. It turns out that the energy (\ref{so5_energy2}) is
minimized by a configuration, where only the strings of length $N_f$
on the first level and strings of length $2N_f$ on the second level
have a finite density (cf. Ref. \cite{Naka86} for the isotropic
case). After inverting the kernels $A^{(1,1)}_{j_{0,1}j_{0,1}}$ and
$A^{(2,2)}_{j_{0,2}j_{0,2}}$ in equation (\ref{so5_densities1}) and
inserting the resulting expression for $\rho^{(1)}_{j_{0,1}}(\lambda)$
and $\rho^{(2)}_{j_{0,2}}(\lambda)$ into the other equations for
$k\neq j_{0,1}$ on the first level and $k\neq j_{0,2}$ on the second
level the integral equations (\ref{so5_densities2}) are found, where
the densities
$\rho^{h(m)}_{j_{0,m}}\leftrightarrow \rho^{(m)}_{j_{0,m}}$ were
redefined and the Fourier-transformed kernels $B^{(l,m)}_{kj}(\omega)$
were introduced:
\begin{equation*}
  \begin{aligned}
    B^{(m,m)}_{j_{0,m}j_{0,m}}&=(-1)^{r^{(m)}_{j_{0,m}}}A^{(\tilde{m},\tilde{m})}_{j_{0,\tilde{m}},j_{0,\tilde{m}}}/K\,,\\
    B^{(m,l)}_{j_{0,m}j_{0,l}}&=(-1)^{r^{(l)}_{j_{0,l}}}A^{(m,l)}_{j_{0,m}j_{0,l}}/K \qquad m\neq l\,,\\
    B^{(l,m)}_{j_{0,l}j}&=(-1)^{l+m}\left(A^{(l,m)}_{j_{0,l}j}A^{(\tilde{l},\tilde{l})}_{j_{0,\tilde{l}}j_{0,\tilde{l}}}-A^{(l,\tilde{l})}_{j_{0,l}j_{0,\tilde{l}}}A^{(\tilde{l},m)}_{j_{0,\tilde{l}}j}\right)/K \qquad j\neq j_{0,l} \text{ if }l=m\,,\\
    B^{(l,m)}_{kj_{0,m}}&=(-1)^{r^{(l)}_k}\left(A^{(l,\tilde{l})}_{kj_{0,\tilde{l}}}B^{(\tilde{l},m)}_{j_{0,\tilde{l}}j_{0,m}}-A^{(l,l)}_{kj_{0,l}}B^{(l,m)}_{j_{0,l}j_{0,m}}\right)\qquad k\neq j_{0,m}\text{ if }l=m\,,\\
    B^{(l,m)}_{kj}&=(-1)^{r^{(l)}_k}\left(A^{(l,\tilde{l})}_{kj_{0,\tilde{l}}}B^{(\tilde{l},m)}_{j_{0,\tilde{l}}j}-A^{(l,l)}_{kj_{0,l}}B^{(l,m)}_{j_{0,l}j}+(-1)^{l+m}A^{(l,m)}_{kj}\right)\qquad k\neq j_{0,l}\,,\,\,j\neq j_{0,m}\,,
  \end{aligned}
\end{equation*}
where $\tilde{m}=m\mod{2}+1$, $\tilde{l}=l\mod{2}+1$ and
\begin{equation}\nonumber
    K(\omega)=A^{(1,1)}_{j_{0,1}j_{0,1}}(\omega)A^{(2,2)}_{j_{0,2}j_{0,2}}(\omega)-A^{(2,1)}_{j_{0,2}j_{0,1}}(\omega)A^{(1,2)}_{j_{0,1}j_{0,2}}(\omega)\,.
\end{equation}
The Fourier-transformed kernels $A^{(m,l)}_{jk}(\omega)$ ($m,l\in\{1,2\}$) can be derived using (\ref{so5_kernelrel1}), (\ref{so5_kernelrel2}), while $A^{(1,1)}_{jk}$ corresponds to $A_{jk}$ in \cite{BoFr18}.

The expressions determining the bare densities $\rho^{(m)}_{0,k}(\lambda)$ of (\ref{so5_baredensity1}), (\ref{so5_baredensity2}) and the bare energies $\epsilon^{(m)}_{0,k}(\lambda)$ of (\ref{so5_energysol1}), (\ref{so5_energysol2}) are
    \begin{equation}
        \begin{aligned}\nonumber
            \rho^{(m)}_{0,k}(\lambda)&=B^{(m,1)}_{kj_{0,1}}\ast\tilde{\rho}^{(1)}_{0,j_{0,1}}(\lambda) \quad \text{for }k=j_{0,m},\\
            \rho^{(m)}_{0,k}(\lambda)&=\delta_{m,1} \tilde{\rho}^{(1)}_{0,k}(\lambda)+  B^{(m,1)}_{kj_{0,1}}\ast\tilde{\rho}^{(1)}_{0,j_{0,1}}(\lambda)\quad \text{for }k\neq j_{0,m}\,,\nonumber\\
            \epsilon^{(m)}_{0,k}(\lambda)&=-B^{(1,m)}_{j_{0,1}k}\ast \tilde{\epsilon}^{(1)}_{0,j_{0,1}}(\lambda)-B^{(2,m)}_{j_{0,2}k}\ast \tilde{\epsilon}^{(2)}_{0,j_{0,2}}(\lambda)\quad \text{for }k= j_{0,m}\,,\nonumber\\
             \epsilon^{(m)}_{0,k}(\lambda)&=\tilde{\epsilon}^{(m)}_{0,k}-B^{(1,m)}_{j_{0,1}k}\ast \tilde{\epsilon}^{(1)}_{0,j_{0,1}}(\lambda)-B^{(2,m)}_{j_{0,2}k}\ast \tilde{\epsilon}^{(2)}_{0,j_{0,2}}(\lambda)\quad \text{for }k\neq j_{0,m}\,.\nonumber
        \end{aligned}
    \end{equation}

\newpage

%

\end{document}